\documentclass[a4paper,twocolumn,10pt,prx,aps,accepted=2026-10-16]{quantumarticle}
\pdfoutput=1
\usepackage[numbers,sort&compress]{natbib}

\usepackage[dvipsnames]{xcolor}
\definecolor{darkred}{rgb}{0.6,0.05,0.05}
\definecolor{darkgreen}{rgb}{0.05,0.6,0.05}
\definecolor{darkblue}{rgb}{0.05,0.05,0.6}
\usepackage[colorlinks]{hyperref}
\hypersetup{colorlinks,
linkcolor=darkred,
filecolor=darkgreen,
urlcolor=darkblue,
citecolor=darkgreen}

\usepackage{mathtools}
\usepackage{amsmath}
\allowdisplaybreaks%
\usepackage{bbold}
\usepackage{amssymb}
\usepackage[normalem]{ulem}
\usepackage{amsthm}
\usepackage{xfrac}
\usepackage{dsfont}
\usepackage{siunitx}
\usepackage{physics}
\usepackage{graphicx}
\usepackage{bm}
\usepackage{layouts}
\usepackage{comment}

\DeclareMathOperator{\vspan}{\mathrm{Span}}
\usepackage{cleveref}

\usepackage{algorithm}
\usepackage{algorithmic}
\makeatletter

\newcommand{\LL}{\mathcal{L}}
\newcommand{\DD}{\mathcal{D}}
\newcommand{\rhot}{\hat{\rho}(t)}
\newcommand{\sss}{\hat{\rho}_{\rm ss}}

\renewcommand{\eqref}[1]{Eq.~(\ref{#1})}

\begin{document}

\author{Andr\'e Melo}
\affiliation{Alice \& Bob, 53 boulevard du G\'en\'eral Martial Valin, 75015 Paris, France}
\author{Gaspard Beugnot}
\affiliation{Alice \& Bob, 53 boulevard du G\'en\'eral Martial Valin, 75015 Paris, France}
\author{Fabrizio Minganti}
\email[E-mail: ]{fabrizio.minganti@gmail.com}
\affiliation{Alice \& Bob, 53 boulevard du G\'en\'eral Martial Valin, 75015 Paris, France}

\title{Variational Perturbation Theory in Open Quantum Systems for Efficient Steady State Computation}

\begin{abstract}
	Determining the steady state of an open quantum system is crucial for characterizing quantum devices and studying various physical phenomena.
	Often, computing a single steady state is insufficient, and it is necessary to explore its dependence on multiple external parameters.
	In such cases, calculating the steady state independently for each combination of parameters quickly becomes intractable.
	Perturbation theory (PT) can mitigate this challenge by expanding steady states around reference parameters, minimizing redundant computations across neighboring parameter values.
	However, PT has two significant limitations: it relies on the pseudo-inverse---a numerically costly operation---and has a limited radius of convergence.
	In this work, we remove both of these roadblocks.
	First, we introduce a variational perturbation theory (VPT) and its multipoint generalization that significantly extends the radius of convergence even in the presence of non-analytic effects such as dissipative phase transitions.
	Then, we develop two numerical strategies that eliminate the need to compute pseudo-inverses.
	The first relies on a single LU decomposition to efficiently construct the steady state within the convergence region, while the second reformulates VPT as a Krylov space recycling problem and uses preconditioned iterative methods.
	We benchmark these approaches across various models, demonstrating their broad applicability and significant improvements over standard PT\@.
\end{abstract}

\maketitle

\section{Introduction}%
\label{Sec:Introduction}

The interaction between a quantum system and its environment gives rise to unique physical phenomena not found in isolated systems~\cite{Haroche_BOOK_Quantum,Wiseman_BOOK_Quantum}.
The Lindblad master equation is a powerful framework for capturing these effects~\cite{BreuerBookOpen, fazio2024manybodyopenquantumsystems} and accurately describes a wide range of experimental platforms, including superconducting circuits~\cite{Blais2021}, polaritons~\cite{Carusotto_RMP_2013_quantum_fluids_light}, atoms~\cite{RevModPhys.85.553}, molecules~\cite{carr2009cold} and ions~\cite{RevModPhys.75.281}.

When studying a Lindblad master equation, the short- or long-time regimes are typically of interest.
During short-time dynamics, the environment introduces small perturbations to the isolated behavior of the system.
Over longer timescales, the system eventually reaches equilibrium with its environment by relaxing towards its steady state.
Mathematically, the steady state is described by a stationary density matrix $\sss$ that satisfies $\LL \sss = 0$, where $\LL$ is the Liouvillian governing the dynamics~\cite{BreuerBookOpen}.
In experimental settings, steady states are often easier to prepare and measure than transient dynamics and provide insight into diverse physical phenomena, including chaos~\cite{ferrari2023steadystatequantumchaosopen,Dahan2022}, phase transitions~\cite{KesslerPRA12,CarmichaelPRX15,MingantPRA18_Spectral}, and other properties such as sub-Poissonian photon statistics~\cite{Hartmann2006}.
Therefore, computing steady states is a central task in the study of open quantum systems.

Although a few models are known to have analytical solutions~\cite{RobertsPRX20,MingantiSciRep16,Drummond_JPA_80_bistability}, most steady states can only be determined numerically.
Open quantum systems, however, are particularly challenging to simulate on classical computers due to the exponential scaling of the Hilbert space and the additional quadratic cost of working with density matrices rather than wave functions.
To address these challenges, various numerical methods have been developed to solve problems in large Hilbert spaces~\cite{WeimerRMP21}, including the quantum trajectory formalism~\cite{plenio1998quantum, daley2014quantum}, renormalization techniques~\cite{FinazziPRL15, RotaPRB17}, tensor~\cite{kshetrimayum2017simple, werner2016positive} and neural network~\cite{PhysRevLett.122.250503, yoshioka2019constructing} ans\"atze, phase space methods~\cite{deuar2007correlations, polkovnikov2010phase}, cluster mean field theory~\cite{JinPRX16}, semidefinite relaxation~\cite{robichon2024bootstrapping} and variational approaches~\cite{WeimerPRL2015}.

Less attention has been devoted to intermediate system sizes, where off-the-shelf sparse linear algebra routines (e.g., $LU$ decomposition) are often sufficient to compute the steady state at a single set of parameters but quickly become inefficient for exploring parameter spaces or mapping phase diagrams.
Additionally, this approach fails to exploit the continuity of the problem, as small changes in parameters typically result in small changes in the steady state.
An alternative strategy relies on perturbation theory for open quantum systems, which extends Rayleigh-Schr\"odinger perturbation theory to this non-Hermitian case~\cite{li2014perturbative,benatti2011asymptotic,li2016resummation}.
However, as we detail below, perturbation theory requires a \emph{computationally expensive} operation known as the \emph{Moore–Penrose pseudo-inverse} and typically has a \emph{limited radius of convergence}.
Beyond perturbative methods, complementary approaches have been proposed to address steady-state computations over extended parameter regions.
In Ref.~\cite{7429-w2mx}, steady states at new parameter values are approximated by projecting onto a reduced basis of solutions computed at selected points in parameter space.
Ref.~\cite{kgsg-3npp} derived an exact expression for the steady state as a function of a single control parameter, yielding a closed-form solution obtained from the unperturbed steady state through an explicit operator inversion.
More broadly, perturbative and hybrid expansion strategies have appeared in other contexts, including structural reanalysis of parametric linear systems~\cite{noor1974approximate} and perturbational--variational treatments of Schr\"odinger eigenvalue problems~\cite{silverman1967perturbational, garrigue2025reducedbasismethodseigenvalue}.
However, these approaches are formulated for static or Hermitian problems and rely on Taylor expansions or energy-based variational principles, and do not directly extend to Liouvillian steady-state problems in open quantum systems.
From a practical perspective, the lack of efficient methods for exploring large parameter regions also impacts experimental efforts.
Indeed, when calibrating quantum hardware, steady state measurements obtained while changing multiple controllable parameters are frequently fitted with numerical simulations~\cite{gebhart2023learning,beaulieu2023observationfirstsecondorderdissipative,berdou2023one}, a task that becomes impractical without scalable computational techniques.

In this work, we introduce \emph{variational perturbation theory} (VPT) for open quantum systems.
We relax the constraints of standard perturbation theory (PT) and recast the perturbative approximation as a low-dimensional variational problem, allowing more efficient computation of steady states across parameter regions and a \emph{wider convergence radius}.
We propose two model-independent methods based on VPT that avoid computing the pseudo-inverse of the Liouvillian, both also enabling the differentiation of the steady state.
First, exploiting the properties of the Liouvillian, we compute the steady state and arbitrarily high orders of the perturbative expansion through a single $LU$ decomposition.
Second, having in mind problems where finding exact matrix factorizations is challenging due to the Hilbert space size, we adapt this approach to construct a variational Krylov space through iterative methods.
We demonstrate VPT's efficiency by applying it to the driven-dissipative Kerr resonator, a two-mode dissipative cat setup, and the dissipative $XYZ$ (Heisenberg) model.
Compared to direct calculations on these models, our approach reduces the computational cost by up to a factor of a hundred.
From a fundamental perspective, we show that efficient generalizations of standard perturbation theory are both possible and practical.

The paper is structured as follows.
We begin in Sec.~\ref{sec:steady_states} with a brief review of how to determine steady states with matrix factorization techniques, followed by an overview of perturbation theory for open quantum systems.
In Sec.~\ref{sec:efficient-computing-recursion}, we present an efficient method for computing the perturbative recurrence relation without relying on a pseudo-inverse.
We then introduce VPT in Sec.~\ref{Sec:VPT} and use it to analyze the driven-dissipative Kerr resonator model in both one- and two-dimensional parameter spaces.
In Sec.~\ref{sec:parameter_estimation}, we demonstrate how to use VPT for efficient parameter estimation and apply it to a dissipative cat qubit model.
Finally, in Sec.~\ref{sec:krylov}, we extend VPT to an iterative Krylov-based approach and apply it to an XYZ model on a $3 \times 3$ lattice.

\section{Open system steady states and how to compute them}%
\label{sec:steady_states}

We consider a generic open quantum system described by a time-independent Lindblad master equation of the form ($\hbar = 1$)
\begin{equation}\label{Eq:Lindblad_ME}
	\partial_t \rhot = \LL \rhot = -i [\hat{H}, \rhot] + \sum_{\mu} \kappa_\mu \DD[\hat{J}_\mu] \rhot,
\end{equation}
where $\hat{H}$ is the Hamiltonian that describes the coherent evolution of the system and $\DD[\hat{J}_\mu]$ are dissipator superoperators acting at a rate $\kappa_\mu$ via the jump operators $\hat{J}_\mu$ as
\begin{equation}\label{Eq:Dissipator_def}
	\DD[\hat{J}_\mu] \rhot = \hat{J}_\mu \rhot \hat{J}_\mu^\dagger -\frac{1}{2} \left(\hat{J}_\mu^\dagger \hat{J}_\mu \rhot +\rhot \hat{J}_\mu^\dagger \hat{J}_\mu\right).
\end{equation}
Here, $\LL$ is the Liouvillian superoperator, generating the Lindblad dynamics.
Upon representing $\LL$ as a matrix, density matrices are mapped to column vectors that we denote with $\ket{\hat{\rho}(t)}$.
We will slightly abuse notation by denoting both the superoperator and its matrix representation as $\LL$: $\LL \rhot$ refers to the superoperator form and $\LL \ket{\rhot}$ to the matrix form.

\subsection{Computing the steady state}
Any open quantum system has at least one steady state
$\sss$, which satisfies $\partial_t \sss =0$ and $\sss = \hat{\rho}(t\to \infty)$~\cite{BreuerBookOpen}.
Equivalently,
\begin{equation}\label{Eq:ss_def}
	\LL \sss =0,
\end{equation}
i.e., the steady states span the kernel of $\LL$.
A common approach to compute $\sss$ is to directly solve~\eqref{Eq:ss_def}.
To avoid converging to the trivial solution $\ket{\sss} = \ket{0, 0, \dots 0}$, the equation for the steady state can be modified as
\begin{equation}\label{Eq:ss_equation_trace}
	\tilde{\LL}\ket{\sss} = (\LL + b \mathcal{T}) \ket{\sss} = \ket{b},
\end{equation}
where
\begin{equation}
	\mathcal{T} \ket{\sss} = \ket{
		\Tr(\sss), 0, \dots, 0
	},
\end{equation}
and $\ket{b} = \ket {b, 0, \dots, 0}$ with $b$ an arbitrary complex number (see Appendix~\ref{Appendix_modified_Liouvillian}).

Equation (\ref{Eq:ss_equation_trace}) is solvable through standard matrix factorization techniques~\cite{trefethen2022numerical}.
Although our results are agnostic to the choice of factorization, we choose $LU$ decomposition due to its numerical efficiency~\footnote{Another common choice is the $QR$ decomposition.
	We use $LU$ since its complexity scales as $\mathcal{O}(2 N^3/3)$ for a system of size $N$ whereas $QR$ scales as $\mathcal{O}(4 N^3/3)$.}.
We get
\begin{equation}%
	\label{eq:lu}
	\tilde{\LL}\ket{\sss} = L\, U \ket{\sss} = \ket{b},
\end{equation}
where $L$ ($U$) is a lower (upper) triangular matrix.
This equation can be solved in two steps (i) solve $ L \ket{\eta} = \ket{b} $ to obtain $\ket{\eta}$ (forward substitution); (ii) solve
$U \ket{\sss} = \ket{\eta}$ to obtain $\ket{\sss}$ (backward substitution).
Since both $L$ and $U$ are triangular matrices, solving these systems requires only $\mathcal{O}(N^2)$ operations for a matrix of size $N \times N$.
More advanced $LU$ decompositions that preserve the sparsity structure also add steps of initial preconditioning on the $\LL$ matrix, such as partial pivoting, further lowering the numerical cost of forward and backward substitutions.

In this work, we are interested in finding the steady state over a wide range of parameters, parameterized by the Liouvillian $\LL(\theta)$ [cf.\ Fig.~\ref{fig:sketch}(a)].
We will benchmark the methods developed below against $LU$ decomposition, whose number of operations to obtain the steady state of a Liouvillian of size $N \times N$ scales as $\mathcal{O}( P \, N^3)$, where $P$ is the number of parameter combinations.

\subsection{Perturbation theory}
\begin{figure*}
	\includegraphics[width=1\linewidth]{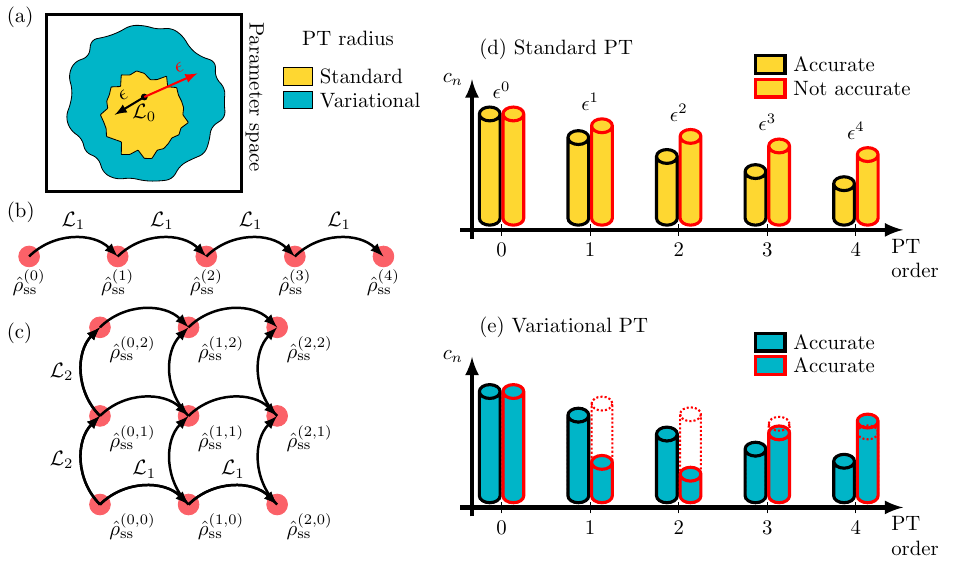}
	\caption{Depiction of the working principle of standard perturbation theory (PT), variational PT (VPT), and of the perturbative recursion relation.
		We are interested in finding steady states of a parameterized Liouvillian $\LL(\varepsilon) = \LL_0 + \varepsilon \LL_1$ for many values of $\varepsilon$ [or, more generally, $\LL(\varepsilon_1, \varepsilon_2, \dots \varepsilon_n) = \LL_0 + \sum \varepsilon_j \LL_{j}$].
		(a) Sketch of the parameter space. At the point $\varepsilon =0$, the steady state $\sss(\varepsilon=0)$ of $\LL(\varepsilon=0) = \LL_0$ is computed. To compute the steady state in neighboring points in the parameter space, we resort to PT and find the set of matrices $\sss^{(n)}$ defined in~\eqref{Eq:recursion_relation}.
		We develop methods that do that efficiently for (b) 1D parameter spaces, (c) 2D parameter spaces and, more generally, for arbitrarily high dimensional parameter space. These methods are based on either re-applying the $LU$ decomposition used to compute $\sss(\varepsilon=0)$, or using approximate methods to build the basis of matrices defining the steady state.
		(d) Having obtained the series of perturbation matrices $\sss^{(n)}$, one can construct the state $\sss(\varepsilon)$ by weighting each matrix according to the power series in~\eqref{Eq:PT_coeff} within the radius of convergence.
		(e) Allowing for a more expressive ansatz in the form of~\eqref{Eq:PT_coeff_general}, one can use $\sss^{(n)}$ to describe the solution on a much wider range of parameters. This makes VPT more efficient than PT, as the ``heavy'' operation of computing the steady state at one point via exact numerical methods becomes less frequent.
	}%
	\label{fig:sketch}
\end{figure*}

Rather than solving the problem for each parameter combination, one can use perturbation theory (PT) to approximate the steady state over contiguous regions in parameter space.
Closely following Refs.~\cite{benatti2011asymptotic, li2014perturbative, li2016resummation}, let us call $\LL_0 = \LL(\bar{\theta})$ for a specific choice $\theta = \bar{\theta}$.
We introduce the Liouvillian in a neighborhood $\varepsilon = \theta - \bar{\theta}$ as
\begin{equation}\label{eq:def-Lepsilon}
	\LL(\varepsilon) = \LL_0 + \varepsilon \, \LL_{1} ,
\end{equation}
where both $\LL_0$ and $\LL_{1}$ are Liouvillians.
PT assumes that the density matrix $\sss$ can be expanded as
\begin{equation}\label{Eq:PT_coeff}
	\sss^{PT}(\varepsilon) = \frac{1}{\mathcal{N}}\sum_{n = 0} ^{\infty} \varepsilon^n \sss^{(n)},
\end{equation}
where $\mathcal{N}$ is a normalization coefficient ensuring $\operatorname{Tr}[\sss^{PT}(\varepsilon)] = 1$.
As discussed in the Appendix~\ref{Appendix_modified_Liouvillian} and sketched in Fig.~\ref{fig:sketch}(b), each term can be obtained through the recurrence relation,
\begin{equation}\label{Eq:recursion_relation}
	\LL_0 \sss^{(0)} =0, \quad  \LL_0 \sss^{(n)} + \LL_1 \sss^{(n-1)} =0.
\end{equation}
Since $\LL_0$ has a nullspace spanned by $\sss$,~\eqref{Eq:recursion_relation} has an infinite number of solutions.
The expansion in~\eqref{Eq:PT_coeff} corresponds to the choice
\begin{equation}\label{Eq:solution_PT}
	\sss^{(n)} = - \LL_0^{\leftharpoonup 1} \LL_1 \sss^{(n-1)},
\end{equation}
where here $\LL_0^{\leftharpoonup 1}$ is the (Moore-Penrose) pseudo-inverse.
Note that this choice of $\sss^{(n)}$ gives the solution with minimum norm and
\begin{equation}\label{Eq:traceless_PT_order_n}
	\braket{\sss^{(0)}}{\sss^{(n)}} =0,
\end{equation}
i.e., the pseudo-inverse removes the nullspace component from $\sss^{(n)}$.
Within the radius of convergence of the series in~\eqref{Eq:PT_coeff}, we can approximate the steady state by truncating the expansion to a desired order $M$.

It is straightforward to generalize~\eqref{Eq:PT_coeff} to multiparameter Liouvillians $\LL(\theta_1, \theta_2, \dots)$.
For instance,
\begin{equation}
	\LL(\varepsilon_1, \varepsilon_2) = \LL_0 + \varepsilon_1 \LL_1 + \varepsilon_2 \LL_2,
\end{equation}
with $\LL_0 = \LL(\bar{\theta}_1, \bar{\theta}_2)$ and $\varepsilon_{1, \, 2} = \theta_{1, \, 2} - \bar{\theta}_{1, \, 2}$, leads to
a two-dimensional perturbative corrections grid  [see Fig.~\ref{fig:sketch}(c)] defined by
\begin{equation}\label{Eq:recursion_relation_2D}
	\begin{split}
		\LL_0 & \sss^{(0, 0)} =0,                                                 \\
		\LL_0 & \sss^{(n, m)} + \LL_1 \sss^{(n-1, m)} + \LL_2 \sss^{(n, m-1)} =0.
	\end{split}
\end{equation}

A crucial limitation of this approach is that constructing the pseudo-inverse is a numerically expensive operation, as it requires diagonalizing $\LL_0$ or computing its singular value decomposition.
We solve this issue in Sec.~\ref{sec:efficient-computing-recursion} and in Sec.~\ref{sec:krylov}.
Furthermore, as we show in Secs.~\ref{Sec:Kerr_1D} and~\ref{Sec:Kerr_2D} and depict in Fig.~\ref{fig:sketch}(d), the convergence radius of the perturbation series is fairly limited especially in the proximity of critical phenomena such as phase transitions.
In Sec.~\ref{Sec:VPT}, we introduce a Variational Perturbation Theory (VPT) that circumvents these limitations.

\section{Efficiently computing the recursive relation}%
\label{sec:efficient-computing-recursion}

Suppose the $LU$ decomposition at a point $\bar{\theta}$ has been computed to obtain $\sss(\bar{\theta})$.
We show how to reuse this factorization to solve the perturbative recurrence relation in~\eqref{Eq:recursion_relation}.
First, we introduce the modified recurrence relation
\begin{equation}%
	\label{Eq:recursion_relation_modified_arbitrary_order_modified}
	\tilde{\LL}_0  \ket{\hat{\rho}^{(0)}} = \ket{b}, \quad
	\tilde{\LL}_0  \ket{\hat{\sigma}^{(n)}} + \LL_1 \ket{\sss^{(n-1)}} =0.
\end{equation}
Crucially, we can reuse $L$ and $U$ in~\eqref{eq:lu} to solve this equation at a marginal numerical cost of $\mathcal{O}(N^2)$.
Since time-evolution is trace-preserving, any Liouvillian admits the identity as a left eigenvector $\bra{\mathbb{1}}$ with zero eigenvalue: $\bra{\mathbb{1}} \LL = 0$.
Applying $\bra{\mathbb 1}$ to~\eqref{Eq:recursion_relation_modified_arbitrary_order_modified} and using $\bra{\mathbb 1}\LL_{0,1}=0$ [see also~\eqref{Eq:trace_super}], we obtain
\begin{equation}
	b\mel{\mathbb{1}}{\mathcal{T}}{\hat{\sigma}^{(n)}} = 0.
\end{equation}
Since $b\neq 0$, this implies $\Tr[\hat\sigma^{(n)}]=0$ and therefore $\mathcal{T} \ket{\hat\sigma^{(n)}}=0$. Substituting back into~\eqref{Eq:recursion_relation_modified_arbitrary_order_modified} eliminates the $b\mathcal T$ contribution to $\tilde{\LL}_0$, leaving
\begin{equation*}
	\LL_0 \ket{\hat{\sigma}^{(n)}} + \LL_1 \ket{\sss^{(n-1)}} = 0.
\end{equation*}
We conclude that, up to a nullspace component, $\ket{\sss^{(n)}}$ and $\ket{\hat{\sigma}^{(n)}}$ are identical.
Using~\eqref{Eq:traceless_PT_order_n} we finally obtain
\begin{equation}
	\ket{\hat{\rho}^{(n)}} = \ket{\hat{\sigma}^{(n)}} -\braket{\hat{\rho}^{(0)}}{\hat{\sigma}^{(n)}} \ket{\hat{\rho}^{(0)}}.
\end{equation}

In summary, although the initial $LU$ factorization of $\tilde{\LL}_0$ is an expensive operation, it enables us to compute the steady state $\sss^{(0)}$ and all perturbative corrections $\sss^{(n)}$ at minimal additional cost without computing the pseudo-inverse of $\LL_0$. Using this approach, the cost to compute the steady state across a phase space of a Liouvillian of size $N \times N$ scales as
\begin{equation}
	\mathcal{O}\left[ \frac{P}{R_{\rm PT}} \left(N^3 + M^d N^2 \right) \right] \simeq \mathcal{O}\left(\frac{P}{R_{\rm PT}} N^3 \right),
\end{equation}
where $R_{\rm PT}$ is the typical convergence volume of the PT series, $P$ is the number of points to compute, and $M$ is the perturbation order. The right-hand side follows in the limit of $M^d \ll N$ and highlights the importance of maximizing the radius of convergence.

\section{Multipoint variational perturbation theory for open quantum systems}%
\label{Sec:VPT}
PT has a fairly limited radius of convergence due to the overly stringent condition on coefficients that leads to~\eqref{Eq:PT_coeff} [see also the discussion in Appendix~\ref{Appendix_PT}] and not because of a lack of expressivity of the vectors $\sss^{(n)}$.
Indeed, they are the Krylov basis of $\LL_0^{\leftharpoonup 1} \LL_1$, and thus form a basis for any finite-dimensional density matrix except in pathological cases.
We thus generalize the series in~\eqref{Eq:PT_coeff} to
\begin{equation}\label{Eq:PT_coeff_general}
	\sss^{PT}(\varepsilon) = \frac{1}{\mathcal{N}}\sum_{n = 0} ^{\infty} c_n(\varepsilon) \sss^{(n)},
\end{equation}
with $\mathcal{N}$  the normalization term.
Within the region where perturbation theory is valid, Eqs.~(\ref{Eq:PT_coeff}) and (\ref{Eq:PT_coeff_general}) must give the same result.
Therefore, a natural choice is to fix $\sss^{(n)}$ to be identical in both expressions.
In turn, this implies that $\sss^{(n)}$ can be efficiently determined using~\eqref{Eq:recursion_relation} and exploiting the strategy developed in Sec.~\ref{sec:efficient-computing-recursion}.
This leaves the task of determining the coefficients $c_n(\varepsilon)$.
While within the radius of convergence of PT we expect $c_n(\varepsilon) \approx \varepsilon^n$ to recover the results of PT, the general form of the coefficients could be determined by assuming $c_n$ to be smooth functions of $\varepsilon$, and then solving
\begin{equation}
	\Big\|\tilde{\LL} (\varepsilon) \sum_{n=0}^{\infty}  c_n(\varepsilon) \ket{\sss^{(n)}}  - \ket{b} \Big\|^2  =0 .
\end{equation}
order-by-order.
This is, however, a cumbersome procedure.
Instead, we adopt a variational approach and approximate $c_n(\varepsilon)$ solving up to order $M$ the least-squares problem
\begin{equation}%
	\label{Eq:least_squares}
	\min_{c_n}\Big\|\tilde{\LL} (\varepsilon) \sum_{n=0}^{M} c_n(\varepsilon) \ket{\sss^{(n)}}  - \ket{b} \Big\|^2 .
\end{equation}
That is, we determine the optimal low-dimensional approximation of the steady state within the span of $\{{\sss^{(0)}}, {\sss^{(1)}}, \dots {\sss^{(M)}}\}$.
The variational nature of~\eqref{Eq:least_squares} ensures that this \textit{variational} PT (VPT) provides a better approximation than standard PT for any finite order $M$ and hence has a larger convergence radius, as we illustrate in Figs.~\ref{fig:sketch}(a) and (e).

As the number of basis elements $M$ increases, $\ket{\sss^{(M)}}$ defined in~\cref{Eq:solution_PT} converges to the top eigenvector of the update operator; this is the power method. Numerically, it is thus convenient to work in a basis obtained by orthonormalizing $\{\ket{\sss^{(0)}}, \ket{\sss^{(1)}}, \dots \ket{\sss^{(M)}}\}$.
Denoting the matrix of orthonormal operators as $\mathcal{Q}$, we rewrite the problem as
\begin{equation}\label{Eq:coeff_minimization_definition}
	\min_{q_n}\| \tilde{\LL}(\varepsilon) \mathcal{Q} \, \vec{q} - \ket{b} \|^2,
\end{equation}
where $\vec{q} = \begin{pmatrix}
		q_0, &
		q_1, &
		\dots
		     &
		q_M
	\end{pmatrix}^T$
and
$q_n$ are the coefficients in the orthonormal basis.

Minimizing~\eqref{Eq:coeff_minimization_definition} is faster than finding the $LU$ decomposition to solve $\LL(\theta) \sss(\theta) =0$, but can still be numerically costly since the matrix $\tilde{\LL}(\varepsilon) \mathcal{Q}$ has size $N \times (M +1)$ for a Liouvillian of size $N \times N$.
To further reduce computational costs we observe that
\begin{equation}%
	\label{eq:bound}
	\begin{split}
		\| \tilde{\LL}(\varepsilon) \mathcal{Q} \, \vec{q}  - \ket{b}\|^2
			                                                                ={} & \|(\mathcal{I} - \mathcal{Q}\mathcal{Q}^\dagger) (\tilde{\LL}(\varepsilon) \mathcal{Q} \, \vec{q}  - \ket{b})\|^2 \\
		                                                                  & {} + \|\mathcal{Q}^\dagger (\tilde{\LL}(\varepsilon) \mathcal{Q} \, \vec{q}  - \ket{b})\|^2 ,
	\end{split}
\end{equation}
where $\mathcal{I}$ is the identity superoperator.
Equation~\eqref{eq:bound} is an equality because $\mathcal{I}-\mathcal{Q}\mathcal{Q}^\dagger$ and $\mathcal{Q}\mathcal{Q}^\dagger$ are orthogonal projectors onto complementary subspaces, and we used $\|\mathcal{Q}\mathcal{Q}^\dagger v\|=\|\mathcal{Q}^\dagger v\|$, which follows from the orthonormality of the columns of $\mathcal{Q}$.
Rather than minimizing the full residual, we choose to null only the second term $\|\mathcal{Q}^\dagger(\tilde{\LL}(\varepsilon)\mathcal{Q}\,\vec{q}-\ket{b})\|$, which vanishes exactly when
\begin{equation}
	\tilde{\LL}_{\rm eff}(\varepsilon) \vec{q}_{\rm eff} =\vec{b}_{\rm eff},
\end{equation}
where $\tilde{\LL}_{\rm eff}(\varepsilon) = \mathcal{Q}^\dagger \tilde{\LL}(\varepsilon) \mathcal{Q}$ has dimension $(M{+}1)\times(M{+}1)$ and $\vec{b}_{\rm eff} = \mathcal{Q}^\dagger\ket{b}$.
Substituting $\vec{q}_{\rm eff}$ into~\eqref{eq:bound}, the residual reduces to its component orthogonal to $\mathrm{range}(\mathcal{Q})$:
\begin{equation}%
	\label{eq:bound_2}
	\begin{multlined}
		\|\tilde{\LL}(\varepsilon)\mathcal{Q}\,\vec{q}_{\rm eff} - \ket{b}\|^2 \\
		= \|(\mathcal{I} - \mathcal{Q}\mathcal{Q}^\dagger) (\tilde{\LL}(\varepsilon)\mathcal{Q}\,\vec{q}_{\rm eff}-\ket{b})\|^2 .
	\end{multlined}
\end{equation}
In the limit in which $\mathcal{Q}$ is a good basis to represent the steady state, this residual is small, and $\vec{q}_{\rm eff}$ provides a good approximation to the solution of~\eqref{Eq:coeff_minimization_definition}.
Notably, since $M$ is typically much smaller than the Hilbert space dimension, solving this reduced equation introduces a significant speedup at the price of finding a slightly suboptimal solution.
In practice, having solved the reduced system for $\vec{q}_{\rm eff}$, we lift the solution back to the full space as $\ket{\sss^{\rm PT}(\varepsilon)} = \mathcal{Q}\vec{q}_{\rm eff}/\mathcal{N}$, where $\mathcal{N}$ is the normalization coefficient ensuring $\operatorname{Tr}[\sss^{\rm PT}(\varepsilon)] = 1$, and directly evaluate the residual of the original problem with a single matrix-vector product against the full Liouvillian:
\begin{equation}
	\|\LL(\varepsilon)\ket{\sss^{\rm PT}(\varepsilon)}\| \;\leq\; {\rm tol}.
\end{equation}
This is a common type of criterion, also used in other variational approaches~\cite{WeimerPRL2015,WeimerRMP21}.
The tolerance should be carefully chosen, especially in the presence of critical phenomena (see Appendix~\ref{Appendix_tolerance}), as lower tolerance implies smaller convergence regions, but greater precision.

Using VPT the cost of computing a phase diagram in a $d$-dimensional parameter space with $P$ points for a $N \times N$ Liouvillian scales as
\begin{equation}
	\begin{split}
		 & \mathcal{O}\left[ \frac{P}{R_{\rm VPT}} \left(N^3 + M^d N^2 \right) + P M^{3d} \right]
		\\
		 & \quad \simeq \mathcal{O}\left(\frac{P}{R_{\rm VPT}} N^3 \right),
	\end{split}
\end{equation}
where $R_{\rm VPT}$ is the typical size of the region where VPT up to order $M$ is valid, and the latter follows under the assumption $M^{d} \ll N$.
Since $R_{\rm VPT} \geq R_{\rm PT}$, using VPT leads to an increase in performance.

\subsection{Variational vs standard perturbation theory in the driven Kerr resonator}

\begin{figure*}[!htpb]
	\includegraphics[width=1\linewidth]{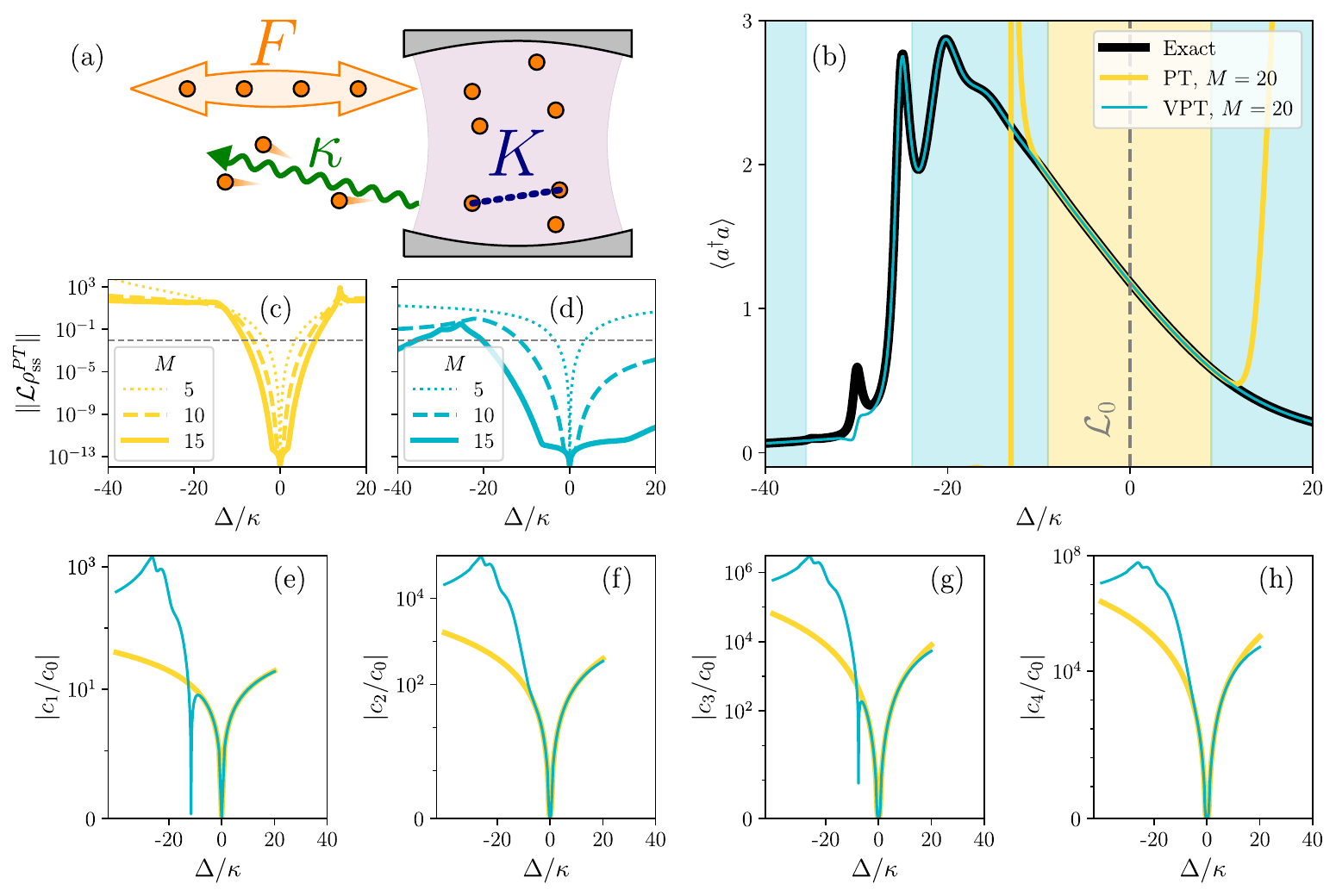}
	\caption{Comparison of standard PT and VPT in the study of the driven-dissipative Kerr resonator described by~\eqref{Eq:Kerr_resonator} and sketched in (a).
		(b) Average photon number $\expval{\hat{a}^\dagger \hat{a}}$ as a function of the detuning $\Delta$  determined using an exact solution (black line), standard PT (yellow line), and VPT (blue line). Both PTs have been performed starting from $\Delta =0$, as indicated by the dashed vertical line. The solid background denotes the region where the perturbative expansion has converged up to $\| \LL(\varepsilon) \sss^{\rm PT}(\varepsilon) \|<10^{-2}$.
		In the yellow region, both standard PT and VPT are convergent.
		In the blue region, only VPT reached convergence.
		(c) Error $\| \LL(\varepsilon) \sss^{\rm PT}(\varepsilon) \|$ as a function of the maximal order of perturbation $M$ for standard PT\@.
		(d) Same as (b), but for VPT\@.
		(e-h) Coefficients obtained by standard [c.f.~\eqref{Eq:PT_coeff}] and variational [c.f.~\eqref{Eq:PT_coeff_general}] PTs.
		Parameters: $K/\kappa = 10$ and $ F/\kappa = 10$.
		Fock space truncation: 30 photons.
	}%
	\label{fig:kerr_resonator}
\end{figure*}

We consider a driven-dissipative Kerr resonator, sketched in Fig.~\ref{fig:kerr_resonator}(a), whose Hamiltonian, in the frame rotating at the pump frequency, reads
\begin{equation}
	\hat{H} = -\Delta \, \hat{a}^\dagger \hat{a} - K/2 \, \hat{a}^\dagger \hat{a}^\dagger \hat{a}  \hat{a} + F \, (\hat{a} + \hat{a}^\dagger).
\end{equation}
Here, $\hat{a}$ ($\hat{a}^\dagger$) is the bosonic annihilation (creation) operator. $\Delta$ is the pump-to-cavity detuning, $K$ is the Kerr nonlinearity, and $F$ is the pump amplitude.
The system is subject to single-photon loss events occurring at a rate $\kappa$, with a Liouvillian reading
\begin{equation}\label{Eq:Kerr_resonator}
	\LL \rhot = - i [\hat{H}, \rhot] + \kappa \, \DD[\hat{a}] \, \rhot.
\end{equation}

\subsubsection{1D example: varying detuning}%
\label{Sec:Kerr_1D}

First, we vary $\Delta$ and fix all other parameters.
We choose $\Delta = 0$ as the initial point for PT and VPT\@. In Fig.~\ref{fig:kerr_resonator}(b) we compare the photon number in the steady state $\expval{\hat{a}^\dagger \hat{a}} = \operatorname{Tr}[\hat{a}^\dagger \hat{a} \sss]$ obtained from the exact solution to the problem~\cite{Drummond_JPA_80_bistability}, standard PT, and VPT\@. The shaded backgrounds indicates the region where $\| \LL \sss^{PT} \|< \, {\rm tol}$ for each PT approach.
We see that VPT recovers the exact solution in a broader region than standard PT, from the vacuum at large detuning, to the coherent-like state at $\Delta \simeq 0$, passing through the multiphoton resonances at intermediate detuning.
Notice also that VPT is capable of finding the solutions in disconnected regions of the parameter space.
In Figs.~\ref{fig:kerr_resonator}(c-d) we investigate the effect of increasing the order $M$ of PT\@. While for standard PT larger $M$ result in a modest gain in precision and in the region of validity [c.f.\ Fig.~\ref{fig:kerr_resonator}(c)], for VPT the region of validity significantly broadens with $M$, as shown in Fig.~\ref{fig:kerr_resonator}(d).
Finally, in Figs.~\ref{fig:kerr_resonator}(e-h) we show the first coefficients of the PT series. As expected, VPT coincides with standard PT close to $\Delta = 0$.

This example highlights the advantages of variational perturbation theory, which will become even more pronounced for larger Hilbert spaces.
Indeed, the effective Liouvillian $\tilde{\LL}_{\rm eff}(\Delta)$ in Fig.~\ref{fig:kerr_resonator}(b) has size $21 \times 21$.
Achieving the same accuracy with a standard truncation of the Hilbert space in the Fock basis requires keeping states up to $\ket{n = 8}$, i.e., a Liouvillian of size $81 \times 81$.

\subsubsection{2D example: simultaneously varying detuning and drive amplitude}%
\label{Sec:Kerr_2D}
\begin{figure}[!htpb]
	\includegraphics[width=\columnwidth]{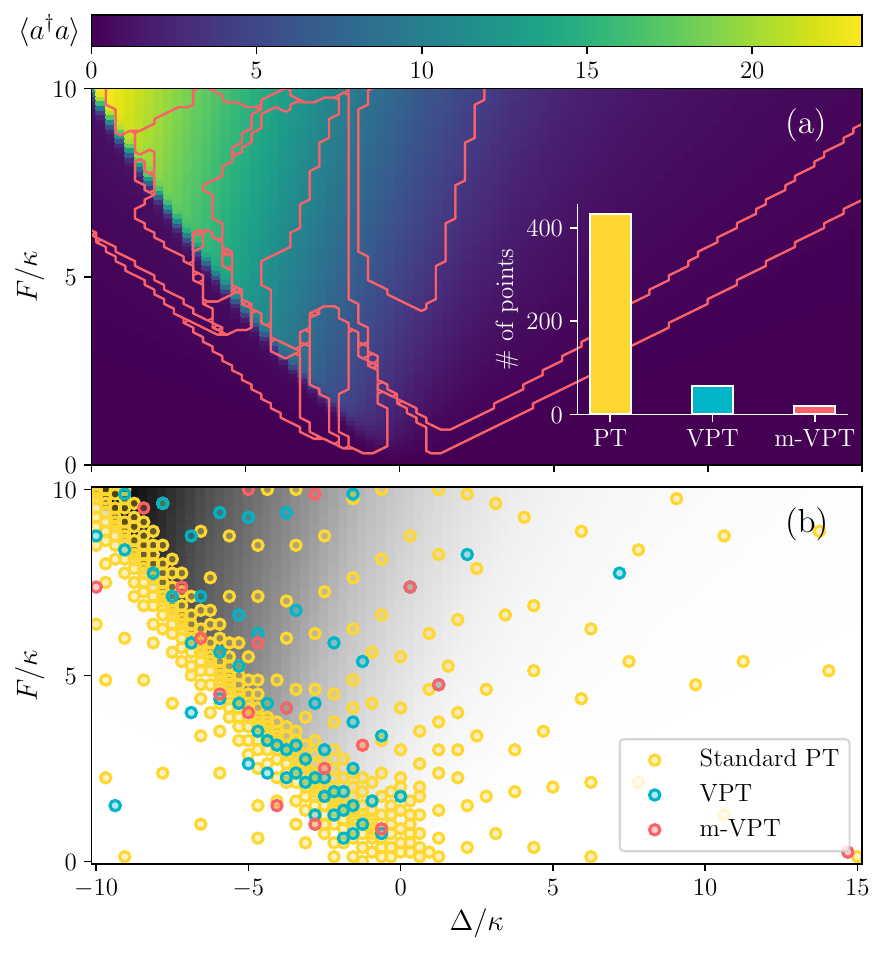}
	\caption{Comparison of the performance of perturbation theory methods to compute the steady state of the driven-dissipative Kerr resonator defined in~\eqref{Eq:Kerr_resonator} in a two-dimensional parameter space. In (a) we plot the average photon number $\langle a^\dagger a\rangle$ in the steady state as a function of detuning $\Delta$ and drive strength $F$ along with the boundaries of the convergence regions of multipoint VPT\@. In (b) we plot the points where we exactly computed the steady state through LU decomposition. At each point we computed a grid of perturbative corrections up to $M = (10, 10)$, totaling 121 perturbation vectors per point, and set the error tolerance to $\| \LL(\varepsilon) \sss^{\rm PT}(\varepsilon) \|<10^{-7}$. Parameters: $K/\kappa = \frac{1}{2}$.
		Fock space truncation: 70 photons.
	}%
	\label{fig:convergence}
\end{figure}
\begin{figure}[!htpb]
	\includegraphics[width=\columnwidth]{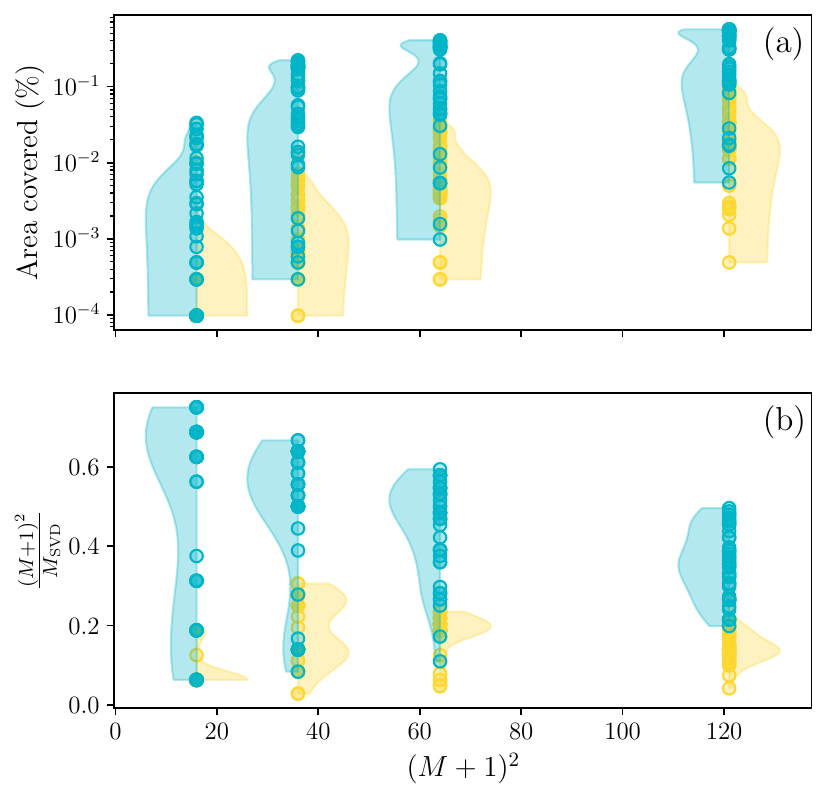}
	\caption{Performance of perturbation theory methods with increasing perturbation order $M$. We study the same system as in Fig. 3 and apply (V)PT at 40 randomly selected points in parameter space. For each resulting convergence region, we compute the optimal low-rank basis (see Appendix~\ref{Appendix:SVD}).
		In both panels, blue violins correspond to VPT and yellow violins to PT, with scatter points indicating individual samples.
		(a) Violin plot of the convergence region area as a function of the VPT basis size $(M+1)^2$.
		(b) Violin plot of $(M+1)^2 / M_\mathrm{SVD}$ as a function of $(M+1)^2$, where $(M+1)^2$ is the number of perturbation vectors used in (V)PT, and $M_\mathrm{SVD}$ is the size of the optimal low-rank basis. A higher ratio indicates a more efficient representation relative to the optimal basis.
		Violin plots illustrate data distributions, with the width representing the density of data points at each value, estimated using a Gaussian kernel.
	}%
	\label{fig:optimality}
\end{figure}

We now apply VPT to compute steady states in a region where $\Delta$ and $F$ change.
Compared to the previous case, we consider a larger value of $\kappa/K$, showcasing the efficiency of VPT in a more dissipative configuration.
We plot the average photon number in the steady state in Fig.~\ref{fig:convergence}(a).
At large positive detuning, the steady state is approximately the vacuum.
Upon decreasing $\Delta$, the system smoothly transitions out of the vacuum phase and gets populated.
At large negative detuning, the system abruptly transitions from a high-photon phase to a vacuum-like one.
As the drive amplitude increases, this transition becomes more pronounced, signaling the emergence of a dissipative phase transition~\cite{BartoloPRA16,CasteelsPRA16,MingantPRA18_Spectral}.
Compared to the previous example, accurately capturing the phase transition will require a significantly smaller tolerance (see the discussion in Appendix~\ref{Appendix_tolerance}).

To cover the parameter space, we first select a random point $(\bar{\Delta}, \bar{F})$ and compute the steady state and its perturbative corrections up to order $M$, introducing the perturbation parameters $\varepsilon_1 = \Delta - \bar{\Delta}$ and $\varepsilon_2= F - \bar{F}$.
Next, we use (V)PT to construct approximate steady states in the neighborhood of the point, ensuring the error remains below the specified tolerance for $\|\LL \sss^{PT}(\varepsilon_1, \varepsilon_2)\|$.
Once we have found the convergence region of (V)PT, we randomly select a new point and repeat the process until the parameter space is completely covered.

In Fig.~\ref{fig:convergence}(b) we compare the performance of PT and VPT\@. Specifically, we plot all the points where we recompute the exact steady state and its perturbative corrections.
We observe that standard PT requires a high density of points near the phase transition region.
This is a consequence of the non-analytic behavior of the system near the critical points where the Liouvillian gap closes.
Strikingly, VPT significantly outperforms PT and requires seven times fewer points to map the whole phase space [cf.\ also the inset in Fig.~\ref{fig:convergence}(a)].
To provide a more quantitative analysis of the performance of VPT and PT, we randomly select 40 points in parameter space where we apply (V)PT and compute the size of the convergence region with increasing perturbation order $M$.
The resulting size distributions, histogrammed in Fig.~\ref{fig:optimality}(a), indicate that VPT consistently yields significantly larger regions of convergence than PT\@. In Fig.~\ref{fig:optimality}(b) we estimate how efficiently (V)PT compresses the steady states by comparing $(M+1)^2$ (the number of perturbation vectors used in (V)PT), to $M_\mathrm{SVD}$---the minimal number of vectors required to span the same convergence region with the same precision as VPT (see details in Appendix~\ref{Appendix:SVD}).
Our results indicate that VPT achieves a compression efficiency close to the theoretical limit.
As the perturbation order $M$ increases, both PT and VPT become less efficient in terms of compression.
However, a higher $M$ also extends the coverage of the parameter space, allowing the method to reach a broader range of solutions.
This reveals an inherent trade-off: increasing the perturbation order enhances parameter space coverage but reduces the optimality of the low-rank representation.

\subsection{Multipoint variational perturbation theory}

The performance of VPT deteriorates in the vicinity of critical points.
To gain intuition into why it breaks down near critical points, we consider a system with a Liouvillian parameterized by a single parameter $\theta$ that transitions at $\theta=0$.
Suppose we computed the steady state and the perturbative series at two points: $\theta_- < 0$ and $\theta_+ > 0$ on the opposite sides of the critical point.
Using standard PT, we have
\begin{equation}
	\sss^{PT, \, \pm}(\varepsilon_\pm) = \sum_n \varepsilon_{\pm}^{n} \sss^{(n)}(\theta_{\pm}),
\end{equation}
where $\varepsilon_{\pm} = (\theta- \theta_{\pm} )$.
None of the states $\sss^{PT, \, \pm}$ will individually describe points near the phase transition, because, roughly speaking, the steady state switches between $\sss(\theta<0)$ and $\sss(\theta>0)$.
The ansatz
\begin{equation}
	\sss(\theta) \approx A(\varepsilon_-) \, \sss^{PT, \, -}(\varepsilon_-) + B(\varepsilon_+) \, \sss^{PT, \, +}(\varepsilon_+),
\end{equation}
instead, can do it, with  $A(\varepsilon_-)$  and $B(\varepsilon_+)$ the coefficients of the left and right steady states, respectively.

Leveraging this intuition, we propose a multipoint variational perturbation theory (m-VPT) based on the perturbative vectors computed at several points:
\begin{equation}
	\sss^{PT}(\theta) = \sum_{n = 0} ^{\infty} \sum_{i} c_n^{(i)}(\theta - \theta_i)\,\sss^{(n)}(\theta_i) .
\end{equation}
We repeat the steady state calculations in Fig.~\ref{fig:convergence} where for each point we consider a basis $\mathcal{Q}$ that contains the local PT correction vectors but also those of the two nearest points.
Compared to VPT, we find that this approach achieves a further three-fold reduction in the number of points where $\sss$ is calculated through $LU$, as shown in the inset of Fig.~\ref{fig:convergence}(a).
In the same panel, we also plot the edges of the convergence regions for this m-VPT, showing they cross the boundary between the high- and low-photon-number phases.

\section{Parameter estimation through efficient gradient computation}%
\label{sec:parameter_estimation}
A common task in operating quantum devices is fitting numerical simulations to experimental results to estimate system parameters.
Here we show how to efficiently use VPT for this task.

Mathematically, we can frame the problem using a Liouvillian $\LL({\phi}, {\theta})$, where ${\phi}$ represents an unknown parameter and ${\theta}$ is the controllable parameter of the experiment (the following equations can be easily generalized to an arbitrary number of fixed and controllable parameters).
We assume that we have access to a set of noisy measurements $\left\{o_{\rm exp}({\theta})\right\}$ associated with the operator $\hat{o}$, whose expectation value $\expval{\hat{o}({\phi}, {\theta})}= \operatorname{Tr}[\sss(\phi, \theta) \hat{o}]$ we can compute numerically.
The objective of a fit routine is then to find ${\phi}^{\rm fit}$ that minimizes the difference between the experimental and numerical data:
\begin{equation}%
	\label{eq:cost}
	{\phi}^{\rm fit} = \min_{{\phi}} \sum_{{\theta}} |\expval{\Delta \hat{o}({\phi}, \theta)}|^2 = \min_{{\phi}} \mathcal{C} ({\phi}),
\end{equation}
where $\expval{\Delta \hat{o}({\phi}, \theta)} = \expval{\hat{o}({\phi}, {\theta})} -  o_{\rm exp}({\theta})$ and $\mathcal{C} ({\phi})$ is the cost function.
One can use gradient-based methods to iteratively find ${\phi}^{\rm fit}$: at each step $n$ of the method, a guess ${\phi}_{n}$ is updated based on $\partial_{{\phi}} \,\mathcal{C}({\phi})$.
Introducing $\delta = \phi - \phi_n$ such that
\begin{equation}
	\LL(\theta, \phi) = \LL(\bar{\theta}, \phi_n) + \varepsilon \LL_1 + \delta \LL_2,
\end{equation}
we get
\begin{equation}
	\begin{split}
		 & \frac{\partial \mathcal{C}({\phi})}{\partial \phi}  \Bigg|_{\phi = \phi_n}  = \frac{\partial \mathcal{C}({\phi})}{\partial \delta} \Bigg|_{\delta = 0} \\
		 & \qquad =     \sum_\theta \expval{\Delta\hat{o}(\phi_n, \theta)}
		\operatorname{Tr}\left[\frac{\partial \sss(\phi_n+\delta, \theta)}{\partial \delta}\Bigg|_{\delta = 0} \hat{o}\right].
	\end{split}
\end{equation}
We thus need to determine $\partial_\delta  \sss(\varepsilon, \delta)|_{\delta = 0}$.

A first way to do this is to assume
\begin{equation}
	\ket{\sss^{PT}(\varepsilon, \delta)} = \frac{1}{\mathcal{N}}  \sum_n c_n(\varepsilon) \left(\sss^{(n, 0)} + \delta  \sss^{(n, 1)} \right).
\end{equation}
That is, we use the VPT ansatz to describe the dependence in $\theta$ and assume that the coefficients $c_n(\varepsilon)$ are independent of $\delta$.
We get
\begin{equation}
	\begin{multlined}
		\frac{\partial \sss(\phi_n+\delta, \theta)}{\partial \delta} \\
		= \frac{ \mathcal{N} \sum_n c_n(\varepsilon) \sss^{(n, 1)} - (\partial_{\delta} \mathcal{N}) \ket{\sss^{PT}(\theta, \delta)}}{\mathcal{N}^2 },
	\end{multlined}
\end{equation}
where
\begin{equation}
	\begin{split}
		\partial_{\delta} \mathcal{N} & = \operatorname{Tr}\left[\sum_n c_n(\varepsilon) \sss^{(n, 1)}\right].
	\end{split}
\end{equation}

A better estimate of the gradient is obtained by allowing all coefficients to simultaneously depend on $\delta$ and $\varepsilon$, namely
\begin{equation}
	\begin{multlined}
		\ket{\sss^{PT}(\varepsilon, \delta)} \\
		= \frac{1}{\mathcal{N}} \left( \sum_n c_{n, 0}(\varepsilon, \delta) \sss^{(n, 0)} + c_{n, 1}(\varepsilon, \delta)  \sss^{(n, 1)} \right).
	\end{multlined}
\end{equation}
Finding the gradient then amounts to solving the following reduced equation, obtained by implicitly differentiating the linear constraint $\tilde{\LL}\ket{\sss}=\ket b$ through the VPT ansatz (see Appendix~\ref{Appendix_gradient} for a derivation, and~\cite{krantz2002implicit} for background on implicit differentiation in numerical analysis):
\begin{equation}\label{Eq:VPT_gradient}
	\begin{split}
		(\LL_0^{\rm eff} + \varepsilon \LL_1^{\rm eff})  \, \frac{\partial \vec{q}(\delta, \varepsilon)}{\partial \delta}\Bigg|_{\delta = 0} = -\LL_2^{\rm eff} \, \vec{q}(0, \varepsilon) \\
		\Rightarrow \frac{\partial \ket{\sss^{PT}(\delta, \varepsilon)}}{\partial \delta}\Bigg|_{\delta = 0} = \mathcal{Q}_{\delta}\frac{\partial \vec{q}(\delta, \varepsilon)}{\partial \delta} \Bigg|_{\delta = 0},
	\end{split}
\end{equation}
where $\mathcal{Q}_\delta$ is the basis obtained by orthonormalizing $\{\sss^{(n,0)}, \sss^{(n,1)}\}$ and $\LL_j^{\rm eff} = \mathcal{Q}_\delta^\dagger \LL_j \mathcal{Q}_\delta$.

\subsection*{Estimating the parameters of a Schr\"odinger cat by fitting buffer spectroscopy measurements}
\begin{figure*}
	\includegraphics[width=1\linewidth]{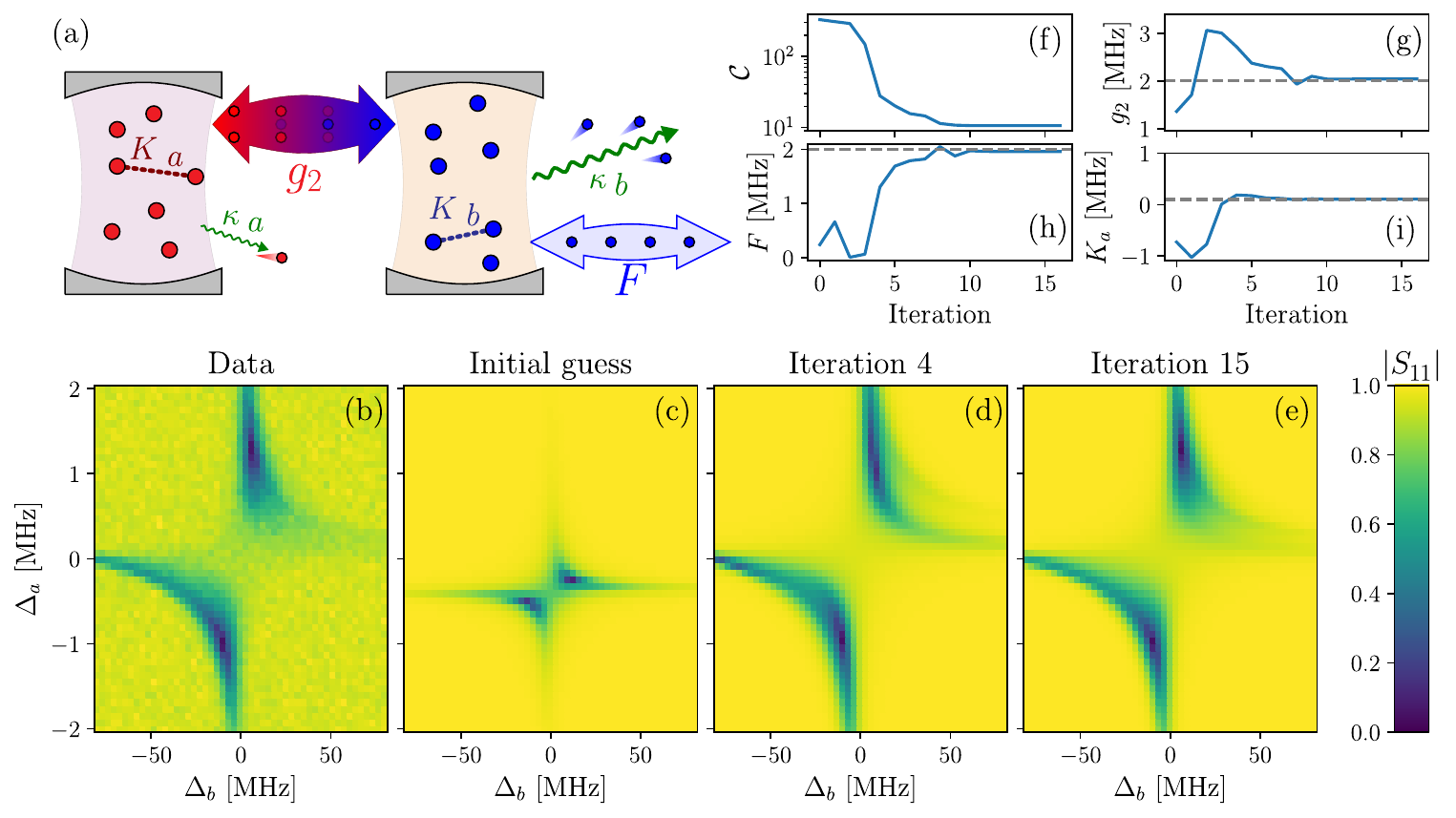}
	\caption{Parameter estimation using VPT and gradient-based optimization for a memory-buffer system sketched in (a) and described by the Liouvillian in~\eqref{Eq:Liouvillian_cat}.
		(b) Synthetic data of the steady state reflection coefficient $S_{11}$ as a function of $\Delta_a$ and $\Delta_b$.
		To mimic experimental imperfections, we add Gaussian noise with average $\mu =0$ and standard deviation $\sigma = 0.02$.
		(c-e) Reflection coefficient at various stages of the optimization.
		The algorithm starts from a random guess of $g_2$, $F$, and $K_a$ and uses the L-BFGS algorithm to optimize them, assuming the remaining parameters are known.
		In (f) we plot the loss as a function of the iteration number, while (g-i) show how the parameters evolve along the optimization process. The dashed lines indicate the true parameters used to generate the data. Parameters (in MHz): $g_2 = 2$, $F = 2$, $K_a = 0.1$, $K_b = 0.3$, $\kappa_a = 0.1$, $\kappa_b = 10$. We fix the Fock space truncation at $10$ photons for mode $\hat{a}$ and $6$ for $\hat{b}$.}%
	\label{fig:spec_fit}
\end{figure*}

We apply VPT and the gradient estimate in~\eqref{Eq:VPT_gradient} on numerically generated data to estimate parameters in a superconducting device that hosts Schrödinger cat states~\cite{berdou2023one}.
The system, which we sketch in Fig.~\ref{fig:spec_fit}(a), consists of a memory mode $\hat{a}$ and a buffer mode $\hat{b}$ that are parametrically coupled through a term $g_2$ that converts a photon in the buffer into two photons in the memory.
The Hamiltonian reads
\begin{equation}%
	\label{eq:cat_hamiltonian}
	\begin{split}
		\hat{H} = & -\Delta_a \hat{a}^\dagger \hat{a} - \Delta_b \hat{b}^\dagger \hat{b} +  g_2 (\hat{a}^2 \hat{b}^{\dag} +{\rm h.c.})
		+  (F \hat{b} +{\rm h.c.})
		\\
		          & \quad- K_a \hat{a}^\dagger \hat{a}^\dagger \hat{a} \hat{a} - K_b \hat{b}^\dagger \hat{b}^\dagger \hat{b} \hat{b} + \chi \hat{a}^\dagger \hat{a}  \hat{b}^\dagger \hat{b}
		,
	\end{split}
\end{equation}
where $\Delta_a$ ($\Delta_b$) is the detuning between the drive and memory (buffer), $K_a$ ($K_b$) is the Kerr nonlinearity, and $F$ is the drive amplitude acting on $\hat{b}$.
Both modes dissipate photons at rates $\kappa_a$ and $\kappa_b$, and the Lindblad master equation reads
\begin{equation}\label{Eq:Liouvillian_cat}
	\LL \rhot = - i [\hat{H}, \rhot] + \kappa_a \DD[\hat{a}] + \kappa_b \DD[\hat{b}] .
\end{equation}
In an ideal configuration to generate cats, $\Delta_a = \Delta_b =0$, and $1/\kappa_b$ is the shortest timescale.
Schr\"odinger cat states persist over times determined by the loss of parity at a rate $\kappa_a \expval{\hat{a}^\dagger \hat{a}}$~\cite{LeghtasScience15,LescanneNatPhys2020}.

Several parameters in~\eqref{eq:cat_hamiltonian} can be directly measured in experiments~\cite{berdou2023one}.
The detuning $\Delta_a$ and $\Delta_b$ are controlled by the relative frequencies of the driving mechanisms.
$K_b$ and $\kappa_b$ can be inferred by probing the $\hat{b}$ mode that has a dedicated measurement and feedline.
Since it is possible to prepare the states $\ket{0}$ or $\ket{1}$ in the memory through the combined action of drives and dissipation, $\chi$ can be inferred by comparing spectroscopy measurements on the $\hat{b}$ mode with the memory in states $\ket{0}$ or $\ket{1}$.
Similarly, $\kappa_a$ can be extracted from the decay rate of a Fock state $\ket{1}$.
The remaining parameters are not straightforward to measure and require more advanced fitting procedures.
Although all these parameters affect the spectroscopic response of the buffer mode, there are no closed or analytical expressions that one can use to fit the system's response.
This motivates us to pursue a simulation-based approach to fit the response of the system and indirectly extract these parameters.

In experiments, a common measurement is the spectroscopic response of the buffer given by $S_{11} = 1 - i \kappa_b\langle \hat{b} \rangle/F$~\cite{berdou2023one}.
First, we pick a set of ground-truth values for $g_2$, $F$, and $K_a$ and numerically generate a dataset of $S_{11}$ as a function of $\Delta_a$ and $\Delta_b$.
To mimic experimental imperfections, we add Gaussian noise to the data, as shown in Fig.~\ref{fig:spec_fit}(b).
Then, starting from an initial random guess of $g_2$, $F$, and $K_a$ and assuming the remaining parameters are known, we try to recover the parameters used in the initial simulation using the scipy implementation~\cite{virtanen2020scipy} of the L-BFGS optimizer~\cite{liu1989limited}. At each step, we minimize the cost function in Eq.~(\ref{eq:cost}) using the VPT gradient estimate provided in~\eqref{Eq:VPT_gradient}.
We show the map of $S_{11}$ computed at various steps of the optimization in Figs.~\ref{fig:spec_fit}(c-e).
As shown in Figs.~\ref{fig:spec_fit}(f-i), we recover the parameters with high accuracy within 15 iterations, although noise introduces small deviations and prevents the loss function from reaching zero.

\section{Variational perturbation theory as a preconditioned Krylov method}%
\label{sec:krylov}
\begin{figure}[!htbp]
	\includegraphics[width=\columnwidth]{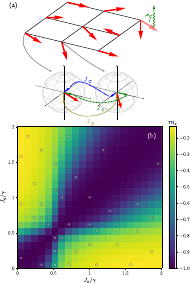}
	\caption{Study of the dissipative XYZ model using preconditioned Krylov methods.
		(a) Pictographic representation of the XYZ model described by~\eqref{Eq:Lindblad_XYZ}.
		Two-level systems interact with their nearest neighbours, with an anisotropic coupling along the three main axes of the Bloch sphere.
		Dissipative events can flip any spin and push it to point in the negative $z$ direction.
		(b) steady state phase diagram of the magnetization $m_z = \expval{\hat{\sigma}_z} $ for a $3 \times 3$ lattice. }%
	\label{fig:xyz}
\end{figure}

Computing exact matrix factorizations of the Liouvillian becomes infeasible for large Hilbert spaces.
An alternative approach is to use iterative procedures
that rely solely on matrix-vector products, such as Krylov-subspace-based methods.
Starting from an initial guess $\ket{q_0}$, these methods iteratively build the Krylov basis of the Liouvillian $\ \{\ket{q_k}\} = \ \{\tilde{\LL}^k \ket{q_{0}}\}$ for $k=0, 1, \dots M$ and search for an approximate solution of~\eqref{Eq:ss_equation_trace} in the form
\begin{equation}\label{eq:def-krylov-subspace-relatedwork}
	\ket{\sss} =  \sum_{k=0}^M c_k \ket{q_k}.
\end{equation}
As a rule of thumb, the closer $\ket{q_0}$ to the actual steady state, the smaller $M$ needs to be.
Common iterative solvers include GMRES~\cite{saad1986gmres} and BiCGSTAB~\cite{van1992bi}, and Conjugate Gradient descent when the linear operator is Hermitian~\cite{hestenes1952methods}.

The convergence rate of these iterative methods depends on the ratio between the largest and smallest singular value of $\tilde{\LL}$, a quantity known as the condition number.
For ill-conditioned problems, one can apply a \emph{preconditioner} $\mathcal{C}$ and solve
\begin{equation}
	\mathcal{C} \tilde{\LL} \ket{\sss} = \mathcal{C} \ket{b}.
\end{equation}
A good preconditioner improves the condition number while being inexpensive to compute and evaluate.
Commonly used preconditioners include band-limited matrices and sparse approximate inverses (see~\cite{saad2003iterative} for a general review and~\cite{nation2015steady} for an in-depth discussion on open quantum systems).
Since the Liouvillian $\tilde{\LL}_0$ is often sparse, a natural choice is to use incomplete $LU$ (iLU), that performs an approximate $LU$ decomposition restricting the outcome to a sparsity pattern closely matching that of $\tilde{\LL}_0$ (i.e., some matrix elements are set to zero).

\subsection{Method and connection with Variational Perturbation Theory}

Our second approach for solving Eq.~(\ref{Eq:ss_equation_trace}) is to use $\tilde{\LL}_0^{\leftharpoonup 1} (\varepsilon=0)$ as a preconditioner and the corresponding steady state $\sss(\varepsilon=0)$ as $\ket{q_0}$.
For a given $M$, we construct the Krylov basis and use it to compute the steady states for the neighboring parameters.
We repeat this procedure on all points where convergence within a fixed tolerance is reached.
Then, we recompute both the  steady state and the preconditioner
$\tilde{\LL}_0^{\leftharpoonup1} (\varepsilon'=0)$.

To understand why $\tilde{\LL}_0^{\leftharpoonup1}$ and $\ket{\sss(\varepsilon=0)}$ are good candidates for preconditioning and Krylov subspaces construction, we note that
\begin{equation}
	\begin{split}
		 & \vspan \left\{\big[\tilde{\LL}_0^{\leftharpoonup1}(\tilde{\LL}_0 + \varepsilon
			                 \LL_1)\big]^k \ket{\sss(\varepsilon=0)} \right\}         \\
		 & \qquad = \vspan \left\{\big(\mathcal{I} + \varepsilon \tilde{\LL}_0^{\leftharpoonup1}
		\LL_1\big)^k \ket{\sss(\varepsilon=0)} \right\}                                          \\
		 & \qquad = \vspan \left\{\big(\tilde{\LL}_0^{\leftharpoonup1} \LL_1\big)^k
		\ket{\sss(\varepsilon=0)} \right\}.
	\end{split}
\end{equation}
Therefore, this preconditioned Krylov method and VPT in~\eqref{Eq:solution_PT} give the same result.
This connects VPT with the larger body of work on recycled Krylov methods, widely used for finite element analysis~\cite{parks2006recycling,soodhalter2014krylov, soodhalter2020survey}. They differ in that VPT recycles the whole subspace, while those methods recycle it portion by portion.

In practice, one rarely has access to $\tilde{\LL}_0^{\leftharpoonup1} (\varepsilon=0)$, and computing it through LU decomposition would defy the purpose of using iterative methods.
We instead compute the iLU decomposition of the Liouvillian that, despite being a worse preconditioner, is significantly cheaper to compute.

Finally, to further speed up the convergence of the Krylov space, one can  \emph{warm start} the algorithm by updating $\ket{q_0}$ using the nearest computed steady state solution.
However, this comes at the cost of re-computing the Krylov subspace.
Despite not being as costly as computing $\LL_0^{-1}$, this still poses a trade-off between the convergence rate and the advantage of reusing a previous Krylov subspace.

\subsection{Phase diagram of the dissipative XYZ model}
The XYZ model describes a set of two-level systems arranged in a square lattice and interacting according to the anisotropic Heisenberg Hamiltonian.
In dimension $D=2$, it reads
\begin{equation}\label{Eq:Hamiltonian_XYZ}
	\hat{H}=\sum_{\langle \vec{n} , \vec{m} \rangle} \left(J_x \hat{\sigma}^x_{\vec{n}} \hat{\sigma}^x_{\vec{m}} + J_y \hat{\sigma}^y_{\vec{n}} \hat{\sigma}^y_{\vec{m}} + J_z \hat{\sigma}^z_{\vec{n}} \hat{\sigma}^z_{\vec{m}}\right)
	,
\end{equation}
where each spin is indexed with a vector $\vec{n} = (n_x, n_y)$, $\hat{\sigma}^\alpha_i$ ($\alpha=x,y,z$) are the Pauli matrices acting on the $\vec{n}$-th site, and the summation includes only nearest neighbor spin pairs $\langle \vec{n} , \vec{m} \rangle$.
The coefficients $J_\alpha$ denote the spin-spin interaction strengths.
Dissipation manifests as incoherent spin flips at a rate $\gamma$ that force spins towards the negative $z$-axis direction, with
\begin{equation}\label{Eq:Lindblad_XYZ}
	\frac{\partial \rhot}{\partial t} =  - i \left[\hat{H}, \rhot\right] +  \gamma \sum_{\vec{n}}  \DD[\hat{\sigma}^-_{\vec{n}}] \rhot \ ,
\end{equation}
where $\hat{\sigma}^\pm_{\vec{n}} = (\hat{\sigma}^x_{\vec{n}}\pm i \hat{\sigma}^y_{\vec{n}})/2$ are the operators raising and lowering the $n$-th spin in the $z$ direction.

The Lindblad master equation (\ref{Eq:Lindblad_XYZ}) remains unchanged under a $\pi$ rotation of all spins about the $z$-axis (transforming $\hat{\sigma}_{\vec{n}}^{x} \to - \hat{\sigma}_{\vec{n}}^{x}$ and $\hat{\sigma}_{\vec{n}}^{y} \to - \hat{\sigma}_{\vec{n}}^{y}$).
This is a $Z_2$ symmetry, meaning that $[\mathcal{P}, \LL] = 0$, with $\mathcal{P} \hat{\rho} = \hat{P}  \hat{\rho}  \hat{P}$ and $ \hat{P} = \prod_{\vec{n}} \hat{\sigma}_{\vec{n}}^{z}$ the parity superoperator and operator, respectively.
In the thermodynamic limit of an infinite lattice and dimension $D\geq2$, the model spontaneously breaks this symmetry and a ferromagnetic order emerges~\cite{LeePRL13,JinPRX16,RotaPRB17,RotaNJP18,HuybrechtsPRB20}.
In this study, we concentrate on a specific parameter range where mean-field theory predicts this transition to occur.

We consider a $3 \times 3$ system with periodic boundary conditions along $x$ and $y$ lattice directions so that the system is translationally invariant.
The Lindblad master equation is then invariant with respect to the transformations $\hat{\sigma}^{\alpha}_{(n_x, n_y)} \to \hat{\sigma}^{\alpha}_{(n_x+1, n_y)}$ and $\hat{\sigma}^{\alpha}_{(n_x, n_y)} \to \hat{\sigma}^{\alpha}_{(n_x, n_y+1)}$.
Defining $\mathcal{T}_x$ and $\mathcal{T}_y$ as superoperators associated with the translational symmetry along the two directions we have $[\mathcal{T}_x, \LL] = [\mathcal{T}_y, \LL] = [\mathcal{P},  \LL] =0$ and $[\mathcal{T}_x, \mathcal{T}_y] = [\mathcal{T}_x, \mathcal{P}] = [\mathcal{T}_y, \mathcal{P}] =0 $.
Let ${\hat\eta_j}$ be a simultaneous eigenbasis of these operators, such that
\begin{equation}
	\mathcal{T}_x \hat{\eta}_j = e^{i \kappa_x^{(j)}} \hat{\eta}_j, \,\mathcal{T}_y \hat{\eta}_j = e^{i \kappa_y^{(j)}} \hat{\eta}_j , \, \mathcal{P} \hat{\eta}_j = e^{i \pi z^{(j)}} \hat{\eta}_j ,
\end{equation}
with $\kappa_x^{(j)}$, $\kappa_y^{(j)}$, and $z^{(j)}$ representing quantum numbers conserved along the dynamics.
For the $3 \times 3$ system considered below, $\kappa_{x,y}^{(j)} \in \{0, 2\pi/3, 4\pi/3\}$ while $z^{(j)} \in \{0, 1\}$.
The Liouvillian written in this basis is block diagonal.
Furthermore, one has that $\mathcal{T}_x \sss = \mathcal{T}_y \sss = \mathcal{P} \sss =  \sss$.
Therefore, the steady state belongs to the symmetry sector with eigenvalues $\kappa_x = \kappa_y =0$ and $z = 0$.

We can thus apply the preconditioned Krylov method to the Liouvillian block containing the steady state.
The full Liouvillian has dimension $2^{18} \approx 2.6 \times 10^{5}$, a size at which repeated $LU$ factorizations across a phase diagram become prohibitively expensive.
Restricting to the symmetry sector containing the steady state reduces this to $\approx 1.5 \times 10^{4}$, where  $LU$ remains tractable but costly, making it a natural target for the iterative approach.
We plot the phase diagram of the magnetization $m_z$ in Fig.~\ref{fig:xyz}(b).
We correctly capture all the expected features, including the transition from the paramagnetic to the ferromagnetic regime.
Similarly to VPT, we observe that the largest concentration of points where the preconditioner is re-computed is in the proximity of the phase transition.

\section{Conclusions}

In this work, we introduced variational perturbation theory (VPT), a generalization of standard perturbation theory for open quantum systems.
We demonstrated its advantages and resilience to detrimental non-analytic behavior across several examples.
To make it numerically viable, we introduced two strategies, one based on exact $LU$ decomposition of the Liouvillian superoperator, and one exploiting incomplete $LU$ to investigate larger system sizes.
Our methods allow rapid exploration of phase diagrams, as well as rapid estimation of steady state gradients, making them ideal tools for parameter fitting routines.

Our method is agnostic to the underlying details of the model under consideration, as we demonstrated.
As such, it can be combined with other techniques such as cluster expansions~\cite{BiellaPRB2018,JinPRX16} and renormalization method~\cite{FinazziPRL15,RotaPRB17} to efficiently investigate more complex problems.
We also plan to further extend VPT to time-domain simulations in systems with time-scale separation where a quasi-steady state regime is present.
Beyond cat states already discussed in this work, typical examples of models with this feature include generic error-corrected quantum systems, metastable configurations, or systems where degrees of freedom can be adiabatically eliminated.

\section*{Author contributions}
A.M.\ and F.M.\ conceived the project and developed the variational perturbation theory and its multipoint generalization.
G.B.\ developed the iterative formulation of the method, establishing its equivalence with preconditioned Krylov-subspace solvers and its relation to subspace recycling.

All authors discussed the results and contributed to writing the manuscript.

Large language models were used to proofread the manuscript and to assist in writing parts of the code implementing the methods presented here. All authors have reviewed and verified the resulting text and code, and take full responsibility for the content of this work.

\acknowledgments{
	We acknowledge support and useful discussion with colleagues at Alice \& Bob. We are grateful to the theoreticians for useful comments, and the experimentalists for providing measurements to test the efficiency of parameter fitting routines.
	In particular, we acknowledge Nicolas Didier, Nathanael Cottet, and their teams.
}

\appendix

\section{Modified Liouvillian}%
\label{Appendix_modified_Liouvillian}

Numerically solving the system of linear equations $\LL \ket{\sss} =0$ outputs the trivial solution $\ket{\sss} = \ket{0, 0, \dots 0}$.
To avoid this issue, one can construct a modified Liouvillian by adding a rank-1 matrix to $\LL$, i.e. $\tilde{\LL} = \LL + \ket b \bra c$, with $b$ and $c$ two vectors.
This operator is full-rank provided that $c$ has non-zero overlap with the kernel of $\LL$.
A natural choice for $c$ is then the identity eigenvector $\bra{\mathbb{1}}$, since $\bra{\mathbb{1}} \LL =0$.
Indeed, any Liouvillian admits identity as a zero left eigenvector because the Lindblad master equation is trace-preserving.
The vector $\ket b$ can be arbitrary; for convenience, in the main text we set it to $\ket{b} = \ket {b, 0, \dots, 0}$ with $b$ an arbitrary complex number.
The equation (\ref{Eq:ss_equation_trace}) follows from this choice, and
\begin{equation}\label{Eq:trace_super}
	\mathcal{T} \ket{\sss} = \ket{1}\bra{\mathbb{1}} \ket{\sss} = \ket{
		\Tr(\sss), 0, \dots, 0
	}.
\end{equation}

\section{Derivation of PT}%
\label{Appendix_PT}

To derive PT in open quantum systems, we expand both the eigenvalues $\lambda_\mu$ and eigenoperators of $\LL = \LL_0 + \varepsilon \LL_1$ as
\begin{equation}
	\lambda_\mu=\sum_{j=0}^{\infty} \varepsilon^j \lambda_\mu^{(j)}, \quad \hat{\rho}_\mu=\sum_{j=0}^{\infty} \varepsilon^j \hat{\rho}_\mu^{(j)},
\end{equation}
where $\lambda_\mu^{(j)}$ and $\hat{\rho}_\mu^{(j)}$ are the $j$-th-order terms for eigenvalues and eigenvectors, respectively.
The recursive relations for $\lambda_\mu^{(j)}$ and $\hat{\rho}_\mu^{(j)}$ can be obtained by expanding
\begin{equation}
	\begin{split}
		\LL(\varepsilon) \hat{\rho}_{\mu}(\varepsilon) & = (\LL_0 + \varepsilon \ \LL_1) \sum_{n = 0} ^{\infty} \varepsilon^n \hat{\rho}_{\mu}^{(n)}                          \\
		                                               & = \sum_{n = 0} ^{\infty} \varepsilon^n \left( \LL_0 \hat{\rho}_{\mu}^{(n)} +  \LL_1  \hat{\rho}_{\mu}^{(n-1)}\right) \\
		                                               & =  \sum_{n=0}^{\infty} \varepsilon^n \lambda_\mu^{(n)}  \sum_{m=0}^{\infty} \varepsilon^m \hat{\rho}_{\mu}^{(m)} .
	\end{split}
\end{equation}
Assuming an order-by-order resolution of this equation we get
\begin{equation}
	\left(\LL_0-\lambda_\mu^{(0)}\right) \hat{\rho}_{\mu}^{(j)}=-\LL_1 \hat{\rho}_{\mu}^{(j-1)} + \sum_{k=1}^j \lambda_\mu^{(k)} \hat{\rho}_{\mu}^{(j-k)}.
\end{equation}
Since we are interested in the steady state $\sss$, $\lambda_\mu^{(0)} =0$.
Furthermore, the identity $\mathbb{1}$ is the left eigenoperator with zero eigenvalue for every Liouvillian and $\bra{\mathbb{1}} \LL_{0, \, 1} =0$.
Thus,
\begin{equation}
	\sum_{k=1}^j \lambda_\mu^{(k)} \braket{\mathbb{1}}{ \hat{\rho}_{\mu}^{(j-k)}} =0,
\end{equation}
for all orders $j$. We conclude that $\lambda_\mu^{(j)} =0$, from which we obtain the recursion relation in~\eqref{Eq:recursion_relation}.

\section{Tolerance criterion}%
\label{Appendix_tolerance}

Tolerance should be fixed with respect to the smallest nonzero Liouvillian eigenvalue.
Indeed, introducing the spectrum of the Liouvillian through the eigenoperators $\hat{\rho}_j$ and eigenvalues $\lambda_j$ defined by $\LL \hat{\rho}_j = \lambda_j \hat{\rho}_j$, we get
\begin{equation}
	\begin{split}
		\sss^{\rm PT}(\varepsilon)                                               ={} & \sum_j c_j \hat{\rho}_j                       \\
		\Longrightarrow \|\LL(\varepsilon) \ket{\sss^{\rm PT}(\varepsilon)}\|^2   ={} & \Big\|\sum_j c_j \lambda_j \ket{\hat{\rho}_j} \Big\|^2 \\
		                                                                        \sim{} & \sum_j |c_j \lambda_j|^2,
	\end{split}
\end{equation}
We conclude that a safe criterion to find the appropriate steady state is $\lambda_1 \gg {\rm tol}$ where $\lambda_1 = \min_{j>0}{|\lambda_j|}$ is the Liouvillian gap, representing the slowest rate in the dynamics of an open quantum system, whose closure indicates the onset of critical phenomena.

\section{Building an optimal low-rank basis using singular value decomposition}%
\label{Appendix:SVD}

To evaluate the optimality of (V)PT's representation, in the main text, we compare it to the best possible low-dimensional approximation of the steady states within the region of validity of the perturbative approximation.
First, we compute the steady states for all parameter combinations. We then stack these state vectors as columns to form a matrix and perform a singular value decomposition (SVD).
The left singular vectors corresponding to the $M_\mathrm{SVD}$ largest singular values form the optimal rank-$M_\mathrm{SVD}$ basis for representing the steady states.
The optimal rank $M_\mathrm{SVD}$ is then chosen to match the accuracy of (V)PT\@.

\section{Efficient computation of the gradient}%
\label{Appendix_gradient}

To determine $\partial_\delta  \sss(\varepsilon, \delta)|_{\delta = 0}$, we first notice that, differentiating~\eqref{Eq:ss_equation_trace}, we get
\begin{equation}
	\partial_\delta [\LL(\phi, \theta) \, \ket{\sss(\phi, \theta)}]\big|_{\delta = 0}  = \partial_{\delta} \ket{b}\big|_{\delta = 0}  = 0 .
\end{equation}
Expanding the derivative, we conclude that the gradient can be computed by solving
\begin{equation}\label{Eq:solution_derivative}
	(\LL_0 + \varepsilon \LL_1)  \, \frac{\partial \ket{\sss(\phi_n+\delta, \theta)}}{\partial \delta}\Bigg|_{\delta = 0} = -\LL_2 \ket{\sss(\phi_n, \theta)}.
\end{equation}
But since $\sss^{PT}(\delta, \varepsilon)$ and $\partial_\delta \ket{\sss^{PT}(\delta, \varepsilon)}$ are approximately spanned by $ \{ \sss^{(n, 0)}, \sss^{(n, 1)} |\, n\in [0, M] \}$,~\eqref{Eq:solution_derivative} can be solved with the same techniques as those used to compute $\vec{q}$.
Namely, we build the orthonormal basis $\mathcal{Q}_{\delta}$ by orthonormalizing the basis of $\{\sss^{(n, 0)}, \sss^{(n, 1)} \}$. Defining
$\LL_j^{\rm eff} = \mathcal{Q}_{\delta}^\dagger \LL_j \mathcal{Q}_{\delta}$, we get the reduced equation
\begin{equation}
	\begin{split}
		(\LL_0^{\rm eff} + \varepsilon \LL_1^{\rm eff})  \, \frac{\partial \vec{q}(\delta, \varepsilon)}{\partial \delta}\Bigg|_{\delta = 0} = -\LL_2^{\rm eff} \, \vec{q}(0, \varepsilon) \\
		\Rightarrow \frac{\partial \ket{\sss^{PT}(\delta, \varepsilon)}}{\partial \delta}\Bigg|_{\delta = 0} =  \mathcal{Q}_{\delta}\frac{\partial \vec{q}(\delta, \varepsilon)}{\partial \delta} \Bigg|_{\delta = 0}
	\end{split}
\end{equation}
Let us also briefly comment on the complexity of computing the gradient for $d$ controllable parameters and $f$ to differentiate. Since we can compute each partial derivative independently along the $f$ parameters, the additional cost of computing the gradient having the phase diagram is
\begin{equation}
	\mathcal{O}\left[ \frac{Pf}{R_{\rm VPT}} \left( M^d N^2 \right) + P f (2 M)^{3d} \right].
\end{equation}

\vspace{2mm}

\bibliographystyle{quantum}
\bibliography{refs}

\begin{thebibliography}{10}

\bibitem{Haroche_BOOK_Quantum}
S.~Haroche and J.~M. Raimond.
\newblock ``Exploring the quantum: Atoms, cavities, and photons''.
\newblock
  \href{https://doi.org/10.1093/acprof:oso/9780198509141.001.0001}{Oxford
  University Press}. Oxford~(2006).

\bibitem{Wiseman_BOOK_Quantum}
H.~M. Wiseman and G.~J. Milburn.
\newblock ``Quantum measurement and control''.
\newblock \href{https://doi.org/10.1017/CBO9780511813948}{Cambridge University
  Press}. Cambridge~(2010).

\bibitem{BreuerBookOpen}
H.~P. Breuer and F.~Petruccione.
\newblock ``The theory of open quantum systems''.
\newblock
  \href{https://doi.org/10.1093/acprof:oso/9780199213900.001.0001}{Oxford
  University Press}. Oxford~(2007).

\bibitem{fazio2024manybodyopenquantumsystems}
R.~Fazio, J.~Keeling, L.~Mazza, and M.~Schir{\'o}.
\newblock ``Many-body open quantum systems''.
\newblock \href{https://doi.org/10.21468/SciPostPhysLectNotes.99}{SciPost Phys.
  Lect. Notes 99}~(2025).

\bibitem{Blais2021}
A.~Blais, A.~L. Grimsmo, S.~M. Girvin, and A.~Wallraff.
\newblock ``Circuit quantum electrodynamics''.
\newblock \href{https://doi.org/10.1103/RevModPhys.93.025005}{Rev. Mod. Phys.
  {\bf 93}, 025005}~(2021).

\bibitem{Carusotto_RMP_2013_quantum_fluids_light}
I.~Carusotto and C.~Ciuti.
\newblock ``Quantum fluids of light''.
\newblock \href{https://doi.org/10.1103/RevModPhys.85.299}{Rev. Mod. Phys. {\bf
  85}, 299--366}~(2013).

\bibitem{RevModPhys.85.553}
H.~Ritsch, P.~Domokos, F.~Brennecke, and T.~Esslinger.
\newblock ``Cold atoms in cavity-generated dynamical optical potentials''.
\newblock \href{https://doi.org/10.1103/RevModPhys.85.553}{Rev. Mod. Phys. {\bf
  85}, 553--601}~(2013).

\bibitem{carr2009cold}
L.~D. Carr, D.~DeMille, R.~V. Krems, and J.~Ye.
\newblock ``Cold and ultracold molecules: science, technology and
  applications''.
\newblock \href{https://doi.org/10.1088/1367-2630/11/5/055049}{New J. Phys.
  {\bf 11}, 055049}~(2009).

\bibitem{RevModPhys.75.281}
D.~Leibfried, R.~Blatt, C.~Monroe, and D.~Wineland.
\newblock ``Quantum dynamics of single trapped ions''.
\newblock \href{https://doi.org/10.1103/RevModPhys.75.281}{Rev. Mod. Phys. {\bf
  75}, 281--324}~(2003).

\bibitem{ferrari2023steadystatequantumchaosopen}
F.~Ferrari, L.~Gravina, D.~Eeltink, P.~Scarlino, V.~Savona, and F.~Minganti.
\newblock ``Dissipative quantum chaos unveiled by stochastic quantum
  trajectories''.
\newblock \href{https://doi.org/10.1103/PhysRevResearch.7.013276}{Phys. Rev.
  Res. {\bf 7}, 013276}~(2025).

\bibitem{Dahan2022}
D.~Dahan, G.~Arwas, and E.~Grosfeld.
\newblock ``Classical and quantum chaos in chirally-driven, dissipative
  {B}ose-{H}ubbard systems''.
\newblock \href{https://doi.org/10.1038/s41534-022-00518-2}{npj Quantum
  Inf.{\bf 8}}~(2022).

\bibitem{KesslerPRA12}
E.~M. Kessler, G.~Giedke, A.~Imamo\v{g}lu, S.~F. Yelin, M.~D. Lukin, and J.~I.
  Cirac.
\newblock ``Dissipative phase transition in a central spin system''.
\newblock \href{https://doi.org/10.1103/PhysRevA.86.012116}{Phys. Rev. A {\bf
  86}, 012116}~(2012).

\bibitem{CarmichaelPRX15}
H.~J. Carmichael.
\newblock ``Breakdown of photon blockade: A dissipative quantum phase
  transition in zero dimensions''.
\newblock \href{https://doi.org/10.1103/PhysRevX.5.031028}{Phys. Rev. X {\bf
  5}, 031028}~(2015).

\bibitem{MingantPRA18_Spectral}
F.~Minganti, A.~Biella, N.~Bartolo, and C.~Ciuti.
\newblock ``Spectral theory of {L}iouvillians for dissipative phase
  transitions''.
\newblock \href{https://doi.org/10.1103/PhysRevA.98.042118}{Phys. Rev. A {\bf
  98}, 042118}~(2018).

\bibitem{Hartmann2006}
M.~J. Hartmann, F.~G. S.~L. Brandão, and M.~B. Plenio.
\newblock ``Strongly interacting polaritons in coupled arrays of cavities''.
\newblock \href{https://doi.org/10.1038/nphys462}{Nat. Phys. {\bf 2},
  849–855}~(2006).

\bibitem{RobertsPRX20}
D.~Roberts and A.~A. Clerk.
\newblock ``Driven-dissipative quantum {K}err resonators: New exact solutions,
  photon blockade and quantum bistability''.
\newblock \href{https://doi.org/10.1103/PhysRevX.10.021022}{Phys. Rev. X {\bf
  10}, 021022}~(2020).

\bibitem{MingantiSciRep16}
F.~Minganti, N.~Bartolo, J.~Lolli, W.~Casteels, and C.~Ciuti.
\newblock ``Exact results for {S}chr{\"o}dinger cats in driven-dissipative
  systems and their feedback control''.
\newblock \href{https://doi.org/10.1038/srep26987}{Sci. Rep. {\bf 6},
  26987}~(2016).

\bibitem{Drummond_JPA_80_bistability}
P.~D. Drummond and D.~F. Walls.
\newblock ``Quantum theory of optical bistability. i. nonlinear polarisability
  model''.
\newblock \href{https://doi.org/10.1088/0305-4470/13/2/034}{J. Phys. A: Math.
  Gen. {\bf 13}, 725--741}~(1980).

\bibitem{WeimerRMP21}
H.~Weimer, A.~Kshetrimayum, and R.~Or\'us.
\newblock ``Simulation methods for open quantum many-body systems''.
\newblock \href{https://doi.org/10.1103/RevModPhys.93.015008}{Rev. Mod. Phys.
  {\bf 93}, 015008}~(2021).

\bibitem{plenio1998quantum}
M.~B. Plenio and P.~L. Knight.
\newblock ``The quantum-jump approach to dissipative dynamics in quantum
  optics''.
\newblock \href{https://doi.org/10.1103/RevModPhys.70.101}{Rev. Mod. Phys. {\bf
  70}, 101}~(1998).

\bibitem{daley2014quantum}
A.~J. Daley.
\newblock ``Quantum trajectories and open many-body quantum systems''.
\newblock \href{https://doi.org/10.1080/00018732.2014.933502}{Adv. Phys. {\bf
  63}, 77--149}~(2014).

\bibitem{FinazziPRL15}
S.~Finazzi, A.~Le~Boit\'e, F.~Storme, A.~Baksic, and C.~Ciuti.
\newblock ``Corner-space renormalization method for driven-dissipative
  two-dimensional correlated systems''.
\newblock \href{https://doi.org/10.1103/PhysRevLett.115.080604}{Phys. Rev.
  Lett. {\bf 115}, 080604}~(2015).

\bibitem{RotaPRB17}
R.~Rota, F.~Storme, N.~Bartolo, R.~Fazio, and C.~Ciuti.
\newblock ``Critical behavior of dissipative two-dimensional spin lattices''.
\newblock \href{https://doi.org/10.1103/PhysRevB.95.134431}{Phys. Rev. B {\bf
  95}, 134431}~(2017).

\bibitem{kshetrimayum2017simple}
A.~Kshetrimayum, H.~Weimer, and R.~Or{\'u}s.
\newblock ``A simple tensor network algorithm for two-dimensional steady
  states''.
\newblock \href{https://doi.org/10.1038/s41467-017-01511-6}{Nat. Commun. {\bf
  8}, 1291}~(2017).

\bibitem{werner2016positive}
A.~H. Werner, D.~Jaschke, P.~Silvi, M.~Kliesch, T.~Calarco, J.~Eisert, and
  S.~Montangero.
\newblock ``Positive tensor network approach for simulating open quantum
  many-body systems''.
\newblock \href{https://doi.org/10.1103/PhysRevLett.116.237201}{Phys. Rev.
  Lett. {\bf 116}, 237201}~(2016).

\bibitem{PhysRevLett.122.250503}
F.~Vicentini, A.~Biella, N.~Regnault, and C.~Ciuti.
\newblock ``Variational neural-network ansatz for steady states in open quantum
  systems''.
\newblock \href{https://doi.org/10.1103/PhysRevLett.122.250503}{Phys. Rev.
  Lett. {\bf 122}, 250503}~(2019).

\bibitem{yoshioka2019constructing}
N.~Yoshioka and R.~Hamazaki.
\newblock ``Constructing neural stationary states for open quantum many-body
  systems''.
\newblock \href{https://doi.org/10.1103/PhysRevB.99.214306}{Phys. Rev. B {\bf
  99}, 214306}~(2019).

\bibitem{deuar2007correlations}
P.~Deuar and P.~D. Drummond.
\newblock ``Correlations in a {BEC} collision: first-principles quantum
  dynamics with 150 000 atoms''.
\newblock \href{https://doi.org/10.1103/PhysRevLett.98.120402}{Phys. Rev. Lett.
  {\bf 98}, 120402}~(2007).

\bibitem{polkovnikov2010phase}
A.~Polkovnikov.
\newblock ``Phase space representation of quantum dynamics''.
\newblock \href{https://doi.org/10.1016/j.aop.2010.02.006}{Ann. Phys. {\bf
  325}, 1790--1852}~(2010).

\bibitem{JinPRX16}
J.~Jin, A.~Biella, O.~Viyuela, L.~Mazza, J.~Keeling, R.~Fazio, and D.~Rossini.
\newblock ``Cluster mean-field approach to the steady-state phase diagram of
  dissipative spin systems''.
\newblock \href{https://doi.org/10.1103/PhysRevX.6.031011}{Phys. Rev. X {\bf
  6}, 031011}~(2016).

\bibitem{robichon2024bootstrapping}
G.~Robichon and A.~Tilloy.
\newblock ``Bootstrapping the stationary state of bosonic open quantum
  systems''.
\newblock \href{https://doi.org/10.1103/5db2-q49l}{Phys. Rev. A {\bf 113},
  062440}~(2026).

\bibitem{WeimerPRL2015}
H.~Weimer.
\newblock ``Variational principle for steady states of dissipative quantum
  many-body systems''.
\newblock \href{https://doi.org/10.1103/PhysRevLett.114.040402}{Phys. Rev.
  Lett. {\bf 114}, 040402}~(2015).

\bibitem{li2014perturbative}
A.~C.~Y. Li, F.~Petruccione, and J.~Koch.
\newblock ``Perturbative approach to {M}arkovian open quantum systems''.
\newblock \href{https://doi.org/10.1038/srep04887}{Sci. Rep. {\bf 4},
  4887}~(2014).

\bibitem{benatti2011asymptotic}
F.~Benatti, A.~Nagy, and H.~Narnhofer.
\newblock ``Asymptotic entanglement and {L}indblad dynamics: a perturbative
  approach''.
\newblock \href{https://doi.org/10.1088/1751-8113/44/15/155303}{J. Phys. A:
  Math. Theor. {\bf 44}, 155303}~(2011).

\bibitem{li2016resummation}
A.~C.~Y. Li, F.~Petruccione, and J.~Koch.
\newblock ``Resummation for nonequilibrium perturbation theory and application
  to open quantum lattices''.
\newblock \href{https://doi.org/10.1103/PhysRevX.6.021037}{Phys. Rev. X {\bf
  6}, 021037}~(2016).

\bibitem{7429-w2mx}
H.~Christiansen, V.~V. Baran, and J.~Paaske.
\newblock ``Reduced basis method for driven-dissipative quantum systems''.
\newblock \href{https://doi.org/10.1103/7429-w2mx}{Phys. Rev. Lett. {\bf 135},
  236503}~(2025).

\bibitem{kgsg-3npp}
O.~Nagib and T.~G. Walker.
\newblock ``Exact steady state of perturbed open quantum systems''.
\newblock \href{https://doi.org/10.1103/kgsg-3npp}{Phys. Rev. Res. {\bf 7},
  033076}~(2025).

\bibitem{noor1974approximate}
A.~K. Noor and H.~E. Lowder.
\newblock ``Approximate techniques of strctural reanalysis''.
\newblock \href{https://doi.org/10.1016/0045-7949(74)90046-7}{Comput. Struct.
  {\bf 4}, 801--812}~(1974).

\bibitem{silverman1967perturbational}
J.~N. Silverman and J.~C. van Leuven.
\newblock ``Perturbational-variational approach to the calculation of
  variational wave functions. i. theory''.
\newblock \href{https://doi.org/10.1103/PhysRev.162.1175}{Phys. Rev. {\bf 162},
  1175}~(1967).

\bibitem{garrigue2025reducedbasismethodseigenvalue}
L.~Garrigue and B.~Stamm.
\newblock ``On reduced basis methods for eigenvalue problems, and on its
  coupling with perturbation theory''~(2025).
\newblock  \href{http://arxiv.org/abs/2408.11924}{arXiv:2408.11924}.

\bibitem{gebhart2023learning}
V.~Gebhart, R.~Santagati, A.~A. Gentile, E.~M. Gauger, D.~Craig, N.~Ares,
  L.~Banchi, F.~Marquardt, L.~Pezze, and C.~Bonato.
\newblock ``Learning quantum systems''.
\newblock \href{https://doi.org/10.1038/s42254-022-00552-1}{Nat. Rev. Phys.
  {\bf 5}, 141--156}~(2023).

\bibitem{beaulieu2023observationfirstsecondorderdissipative}
G.~Beaulieu, F.~Minganti, S.~Frasca, V.~Savona, S.~Felicetti, R.~Di~Candia, and
  P.~Scarlino.
\newblock ``Observation of first- and second-order dissipative phase
  transitions in a two-photon driven {K}err resonator''.
\newblock \href{https://doi.org/10.1038/s41467-025-56830-w}{Nat. Commun. {\bf
  16}, 1951}~(2025).

\bibitem{berdou2023one}
C.~Berdou, A.~Murani, U.~Reglade, W.~C. Smith, M.~Villiers, J.~Palomo,
  M.~Rosticher, A.~Denis, P.~Morfin, M.~Delbecq, et~al.
\newblock ``One hundred second bit-flip time in a two-photon dissipative
  oscillator''.
\newblock \href{https://doi.org/10.1103/PRXQuantum.4.020350}{PRX Quantum {\bf
  4}, 020350}~(2023).

\bibitem{trefethen2022numerical}
L.~N. Trefethen and D.~Bau.
\newblock ``Numerical linear algebra''.
\newblock \href{https://doi.org/10.1137/1.9780898719574}{SIAM}. ~(1997).

\bibitem{BartoloPRA16}
N.~Bartolo, F.~Minganti, W.~Casteels, and C.~Ciuti.
\newblock ``Exact steady state of a {{K}err} resonator with one- and two-photon
  driving and dissipation: Controllable {W}igner-function multimodality and
  dissipative phase transitions''.
\newblock \href{https://doi.org/10.1103/PhysRevA.94.033841}{Phys. Rev. A {\bf
  94}, 033841}~(2016).

\bibitem{CasteelsPRA16}
W.~Casteels, F.~Storme, A.~Le~Boit\'e, and C.~Ciuti.
\newblock ``Power laws in the dynamic hysteresis of quantum nonlinear photonic
  resonators''.
\newblock \href{https://doi.org/10.1103/PhysRevA.93.033824}{Phys. Rev. A {\bf
  93}, 033824}~(2016).

\bibitem{krantz2002implicit}
S.~G. Krantz and H.~R. Parks.
\newblock ``The implicit function theorem: history, theory, and applications''.
\newblock \href{https://doi.org/10.1007/978-1-4612-0059-8}{Springer}. ~(2002).

\bibitem{LeghtasScience15}
Z.~Leghtas, S.~Touzard, I.~M. Pop, A.~Kou, B.~Vlastakis, A.~Petrenko, K.~M.
  Sliwa, A.~Narla, S.~Shankar, M.~J. Hatridge, M.~Reagor, L.~Frunzio, R.~J.
  Schoelkopf, M.~Mirrahimi, and M.~H. Devoret.
\newblock ``Confining the state of light to a quantum manifold by engineered
  two-photon loss''.
\newblock \href{https://doi.org/10.1126/science.aaa2085}{Science {\bf 347},
  853--857}~(2015).

\bibitem{LescanneNatPhys2020}
R.~Lescanne, M.~Villiers, T.~Peronnin, A.~Sarlette, M.~Delbecq, B.~Huard,
  T.~Kontos, M.~Mirrahimi, and Z.~Leghtas.
\newblock ``Exponential suppression of bit-flips in a qubit encoded in an
  oscillator''.
\newblock \href{https://doi.org/10.1038/s41567-020-0824-x}{Nat. Phys. {\bf 16},
  509--513}~(2020).

\bibitem{virtanen2020scipy}
P.~Virtanen, R.~Gommers, T.~E. Oliphant, M.~Haberland, T.~Reddy, D.~Cournapeau,
  E.~Burovski, P.~Peterson, W.~Weckesser, J.~Bright, et~al.
\newblock ``{S}ci{P}y 1.0: fundamental algorithms for scientific computing in
  {P}ython''.
\newblock \href{https://doi.org/10.1038/s41592-019-0686-2}{Nat. Methods {\bf
  17}, 261--272}~(2020).

\bibitem{liu1989limited}
D.~C. Liu and J.~Nocedal.
\newblock ``On the limited memory {BFGS} method for large scale optimization''.
\newblock \href{https://doi.org/10.1007/bf01589116}{Math. Program. {\bf 45},
  503--528}~(1989).

\bibitem{saad1986gmres}
Y.~Saad and M.~H. Schultz.
\newblock ``{GMRES}: A generalized minimal residual algorithm for solving
  nonsymmetric linear systems''.
\newblock \href{https://doi.org/10.1137/0907058}{SIAM J. Sci. Stat. Comput.
  {\bf 7}, 856--869}~(1986).

\bibitem{van1992bi}
H.~A. Van~der Vorst.
\newblock ``{{Bi-CG}STAB}: A fast and smoothly converging variant of {Bi-CG}
  for the solution of nonsymmetric linear systems''.
\newblock \href{https://doi.org/10.1137/0913035}{SIAM J. Sci. Stat. Comput.
  {\bf 13}, 631--644}~(1992).

\bibitem{hestenes1952methods}
M.~R. Hestenes, E.~Stiefel, et~al.
\newblock ``Methods of conjugate gradients for solving linear systems''.
\newblock \href{https://doi.org/10.6028/jres.049.044}{J. Res. Natl. Bur. Stand.
  {\bf 49}, 409}~(1952).

\bibitem{saad2003iterative}
Y.~Saad.
\newblock ``Iterative methods for sparse linear systems''.
\newblock \href{https://doi.org/10.1137/1.9780898718003}{SIAM}. ~(2003).

\bibitem{nation2015steady}
P.~D. Nation.
\newblock ``Steady-state solution methods for open quantum optical
  systems''~(2015).
\newblock  \href{http://arxiv.org/abs/1504.06768}{arXiv:1504.06768}.

\bibitem{parks2006recycling}
M.~L. Parks, E.~de~Sturler, G.~Mackey, D.~D. Johnson, and S.~Maiti.
\newblock ``Recycling {K}rylov subspaces for sequences of linear systems''.
\newblock \href{https://doi.org/10.1137/040607277}{SIAM J. Sci. Comput. {\bf
  28}, 1651--1674}~(2006).

\bibitem{soodhalter2014krylov}
K.~M. Soodhalter, D.~B. Szyld, and F.~Xue.
\newblock ``{K}rylov subspace recycling for sequences of shifted linear
  systems''.
\newblock \href{https://doi.org/10.1016/j.apnum.2014.02.006}{Appl. Numer. Math.
  {\bf 81}, 105--118}~(2014).

\bibitem{soodhalter2020survey}
K.~M. Soodhalter, E.~de~Sturler, and M.~E. Kilmer.
\newblock ``A survey of subspace recycling iterative methods''.
\newblock \href{https://doi.org/10.1002/gamm.202000016}{GAMM-Mitt. {\bf 43},
  e202000016}~(2020).

\bibitem{LeePRL13}
T.~E. Lee, S.~Gopalakrishnan, and M.~D. Lukin.
\newblock ``Unconventional magnetism via optical pumping of interacting spin
  systems''.
\newblock \href{https://doi.org/10.1103/PhysRevLett.110.257204}{Phys. Rev.
  Lett. {\bf 110}, 257204}~(2013).

\bibitem{RotaNJP18}
R.~Rota, F.~Minganti, A.~Biella, and C.~Ciuti.
\newblock ``Dynamical properties of dissipative {${XYZ}$} {H}eisenberg
  lattices''.
\newblock \href{https://doi.org/10.1088/1367-2630/aab703}{New J. Phys. {\bf
  20}, 045003}~(2018).

\bibitem{HuybrechtsPRB20}
D.~Huybrechts, F.~Minganti, F.~Nori, M.~Wouters, and N.~Shammah.
\newblock ``Validity of mean-field theory in a dissipative critical system:
  {L}iouvillian gap, $\mathbb{PT}$-symmetric antigap, and permutational
  symmetry in the ${XYZ}$ model''.
\newblock \href{https://doi.org/10.1103/PhysRevB.101.214302}{Phys. Rev. B {\bf
  101}, 214302}~(2020).

\bibitem{BiellaPRB2018}
A.~Biella, J.~Jin, O.~Viyuela, C.~Ciuti, R.~Fazio, and D.~Rossini.
\newblock ``Linked cluster expansions for open quantum systems on a lattice''.
\newblock \href{https://doi.org/10.1103/PhysRevB.97.035103}{Phys. Rev. B {\bf
  97}, 035103}~(2018).

\end{thebibliography}

\end{document}